\documentclass[sigconf, authorversion]{acmart}

\usepackage{enumitem}
\usepackage{graphicx}
\usepackage{balance}
\usepackage{subcaption}
\usepackage{amsmath}
\usepackage{color}
\usepackage{xcolor}
\usepackage{booktabs}
\usepackage{multirow}
\usepackage{framed}
\usepackage{fancyhdr}
\usepackage{setspace}
\usepackage{acmart-taps}
\definecolor{TFFrameColor}{rgb}{0,0,0}

\AtBeginDocument{%
  }

\copyrightyear{2025}
\acmYear{2025}
\setcopyright{cc}
\setcctype{by}
\acmConference[CHI '25]{CHI Conference on Human Factors in Computing Systems}{April 26-May 1, 2025}{Yokohama, Japan}
\acmBooktitle{CHI Conference on Human Factors in Computing Systems (CHI '25), April 26-May 1, 2025, Yokohama, Japan}
\acmDOI{10.1145/3706598.3714117}
\acmISBN{979-8-4007-1394-1/25/04}

\begin{document}

\title[The Hall of AI Fears and Hopes]{The Hall of AI Fears and Hopes: Comparing the Views\\of AI Influencers and those of Members of the U.S. Public\\Through an Interactive Platform}

\author{Gustavo Moreira}
\orcid{0000-0002-4762-7703}
\affiliation{
  \institution{University of Illinois Chicago}
  \city{Chicago}
  \country{USA}}
\email{gmorei3@uic.edu}

\author{Edyta Paulina Bogucka}
\orcid{0000-0002-8774-2386}
\affiliation{
  \institution{Nokia Bell Labs}
  \city{Cambridge}
  \country{United Kingdom}}
\affiliation{
  \institution{University of Cambridge}
  \city{Cambridge}
  \country{United Kingdom}}
\email{edyta.bogucka@nokia-bell-labs.com}

\author{Marios Constantinides}
\orcid{0000-0003-1454-0641}
\affiliation{
  \institution{CYENS Centre of Excellence}
  \city{Nicosia}
  \country{Cyprus}}
\affiliation{
  \institution{Nokia Bell Labs}
  \city{Cambridge}
  \country{United Kingdom}}
\email{marios.constantinides@cyens.org.cy}

\author{Daniele Quercia}
\orcid{0000-0001-9461-5804}
\affiliation{
  \institution{Nokia Bell Labs}
  \city{Cambridge}
  \country{United Kingdom}}
\affiliation{
  \institution{Politecnico di Torino}
  \city{Turin}
  \country{Italy}}
\email{quercia@cantab.net}

\renewcommand{\shortauthors}{Moreira et al.}

\begin{abstract}
AI development is shaped by academics and industry leaders---let us call them ``influencers''---but it is unclear how their views align with those of the public. To address this gap, we developed an interactive platform that served as a data collection tool for exploring public views on AI, including their fears, hopes, and overall sense of hopefulness. We made the platform available to 330 participants representative of the U.S. population in terms of age, sex, ethnicity, and political leaning, and compared their views with those of 100 AI influencers identified by Time magazine. The public fears AI getting out of control, while influencers emphasize regulation, seemingly to deflect attention from their alleged focus on monetizing AI's potential. Interestingly, the views of AI influencers from underrepresented groups such as women and people of color often differ from the views of underrepresented groups in the public.
\end{abstract}

\ccsdesc[500]{Human-centered computing~Empirical studies in HCI}
\ccsdesc[500]{Human-centered computing~Empirical studies in collaborative and social computing}
\ccsdesc[500]{Human-centered computing~HCI design and evaluation methods}
\ccsdesc[500]{Human-centered computing~Visualization systems and tools}
\ccsdesc[300]{Computing methodologies~Artificial intelligence}
\ccsdesc[300]{Social and professional topics~Codes of ethics}
\ccsdesc[300]{Information systems~Crowdsourcing}

\keywords{responsible AI, ethical AI, AI governance, empirical ethics, value alignment, AI fears, AI hopes, AI influencers, participatory AI ethics, crowdsourcing}

\begin{teaserfigure}
  \centering
  \includegraphics[width=0.96\textwidth]{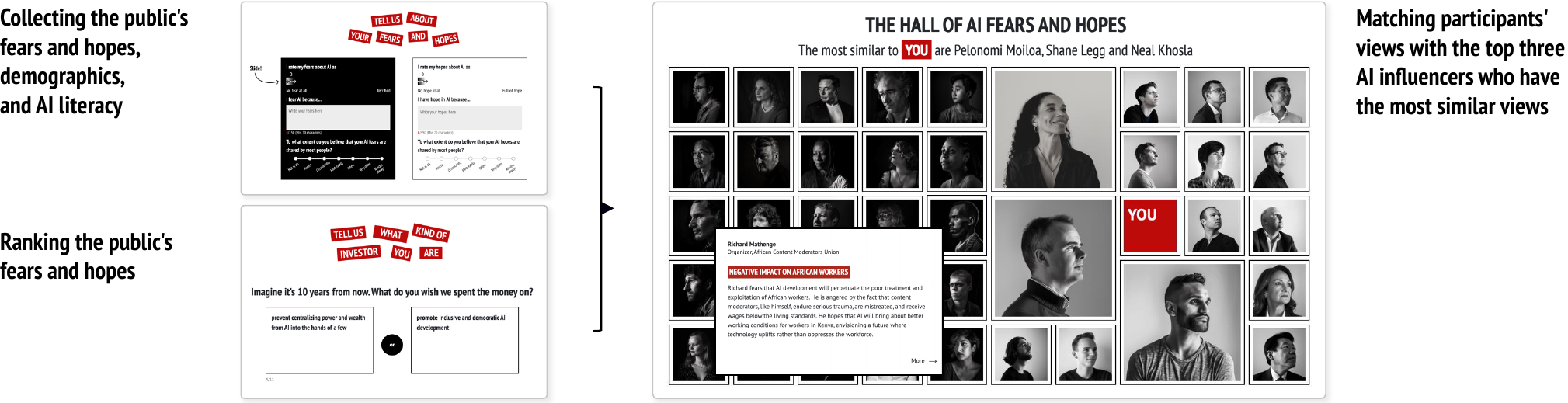}
  \caption{``The Hall of AI Fears and Hopes'' is an interactive platform for collecting the public's views on AI.}
  \Description{A teaser figure showing three sections of ``The Hall of AI Fears and Hopes'' – our interactive platform for collecting the public's views on AI. The first screen contains a questionnaire to gather their fears, hopes, and overall sense of hopefulness. The second screen contains a pairwise comparison game, where participants are asked to imagine a future 10 years from now and answer “What do you wish we spent the money on?” Then, they need to choose between two cards showing randomly 10 pairs of fears and hopes of AI influencers. The third screen presents a gallery of black-and-white portraits of AI influencers. It uses the previously collected data to match participants' views with the top three AI influencers who have the most similar views. The portraits of these influencers are enlarged and moved to the top of the gallery.}
  \label{fig:teaser}
\end{teaserfigure}

\maketitle
\section{Introduction}
\label{sec:introduction}

The rapid advancement of Artificial Intelligence (AI) systems has been shaped by individuals commonly referred to as ``AI influencers''--researchers, technologists, policymakers, and thought leaders--whose expertise, advocacy, or commercial interests shape public discourse, policy, and industry trends~\cite{corbett2023power, delgado2023participatory}. Researchers, for example, push the boundaries of what is possible with AI by shaping the theoretical underpinnings that guide its development~\cite{impactRAI2024}. On the other hand, industry leaders and affluent individuals enable the commercialization and widespread adoption of these technologies through their financial and organizational resources \cite{patentAnalysis2020}. By investing in AI startups, funding research initiatives, and shaping market trends, these individuals ensure that cutting-edge AI technologies transition from academic laboratories to real-world applications, making them accessible to the public~\cite{researchOnIndustryAIPractices2024}. Given the considerable power these individuals hold, they often emerge as key decision-makers in the AI landscape \cite{organisationalPractices2021, AIStakeholders2022}. The decisions they make are not isolated; they carry profound implications for society at large. These decisions can then determine the trajectory of AI development, influence regulatory frameworks \cite{ai_standard_influencers_2025}, and shape public perception and trust in AI systems \cite{kelley2021exciting}. Consequently, the impact of these decisions is far-reaching, affecting not only the technological landscape but also societal structures, and the daily lives of the public.

However, the public is frequently excluded from these critical decision-making processes~\cite{birhane2022power, corbett2023power}. Despite being the most affected by AI technologies (yet relegated to the role of ``AI subjects'' \cite{EUACT2024} or ``human subjects'' \cite{nist2023aiRisk}), the public's voice is often marginalized or entirely absent in discussions about AI development and governance \cite{ojewale2024towards}. When policies are developed without adequately reflecting public concerns, the public may be compelled to follow rules they find unfair or misaligned with their values. This issue is worsened by the fact that many key decisions around AI are made by unelected officials, corporate executives, and industry leaders rather than through democratic processes~\cite{schaake2024tech}. Similar observations were made on smart cities wherein citizens are either represented by professional and bureaucratic elites, or are entirely absent from key decision making~\cite{shelton2019actually, whitney2021hci}. Concentrating decision-making power beyond democratic oversight risks alienating the public and eroding trust in AI and its governance \cite{awad2020approach, Moss2021assemblingAccountability, AITerminology_2022}.

In particular, work in the Human-Computer Interaction (HCI) community has demonstrated the importance of including diverse public viewpoints in AI development and governance \cite{jakesch2022different, AIRoles_2023, situateAIGuidebook_2024}. A thematic analysis of public views has identified several key drivers of AI-phobia: AI substitutability, AI accountability, AI literacy, and AI fever, all of which point out public anxiety surrounding AI's impact on employment and healthcare~\cite{AIPhobia2023, nader2024public}. These findings reveal that concerns over job displacement and the fairness of AI's integration into healthcare are top priorities for the public~\cite{cave2019scary, googleAIsurvey}, reinforcing the need for a more inclusive AI governance \cite{AIContestability_2024}.

A growing body of literature has focused on understanding public views toward AI in isolation, without adequately examining how these views compare to those of the influential figures who shape the AI discourse~\cite{sartori2023minding, kelley2021exciting, huang2024collective, scantamburlo2023artificial, cave2019scary, kieslich2024regulating}. This lack of comparative analysis creates a gap in our understanding: \emph{we do not fully comprehend whether the decisions made by AI influencers reflect the desires and concerns of the public}. This (mis)alignment could either exacerbate existing societal inequalities or foster more equitable and inclusive AI systems. Recent studies from the HCI community stress the need to bridge this gap by creating tools and processes that ensure public concerns are not only heard but actively inform AI policy and development \cite{methodsForCollectingAIOpinions2023, situateAIGuidebook_2024, AIContestability_2024}. Therefore, understanding this dynamic is essential for developing AI technologies that are not only innovative but also socially responsible and aligned with the public's best interests. To investigate this dynamic, we ask:

\begin{itemize}
    \item[(RQ\textsubscript{1})] What are the views of a sample of the U.S. public (called ``the public'' in the rest of the paper) on AI?
    \item[(RQ\textsubscript{2})] What are the views of AI influencers?
    \item[(RQ\textsubscript{3})] Which subgroups of AI influencers have views that are most (or least) aligned with our sample of the public?
    \item[(RQ\textsubscript{4})] What themes emerge from the most (or least) aligned views?
\end{itemize}

To address these questions, we conducted a three-phase study. First, we curated a dataset of views from 100 influential figures in AI, as identified by Time magazine, including researchers, innovators, and policymakers. Second, we co-designed an interactive platform, ``The Hall of AI Fears and Hopes'', with two designers, and tested it with 30 participants to refine the data collection process. We deployed this platform on Prolific with 330 U.S. participants, representative in terms of age, sex, ethnicity, and political leaning, to collect their views. Third, we conducted a comparative analysis to identify patterns of alignment and misalignment between public views and those of AI influencers. In so doing, our study makes two main contributions:

\begin{enumerate}
\item We proposed a structured approach to compare how AI influencers and the public perceive AI, including (1) a dataset of views from influential AI figures (\S\ref{subsec:method1}); (2) a platform that allows people to explore these influencer views and share their own (\S\ref{subsec:method2}); and (3) metrics and methods to analyze the differences between these views (\S\ref{subsec:method3}).

\item We provided empirical insights into (mis)alignment patterns between AI influencers and the public. Our analysis reveals that older influencers shared fewer views with the public, while younger influencers, academics, and non-billionaires aligned more closely with public opinions. Interestingly, AI influencers from underrepresented groups (e.g., women, people of color) showed poor alignment with the same underrepresented groups within the public sample, indicating a disconnect even among historically underrepresented voices. At the same time, each group has different concerns: the public worries about losing control to AI and perceives only a few benefits, while influencers focus on supposedly controlling AI by calling for regulations and see many potential benefits of AI.
\end{enumerate}

Building on these contributions, we discuss how our results align with or challenge existing literature and highlight the need for follow-up studies across different contexts and a broader range of AI influencers (\S\ref{sec:discussion}). While our analysis provides a valuable starting point with a subset of 100 influencers, it may not fully capture the diversity of views within the AI community. Therefore, we propose exploring alternative sampling approaches to better capture this diversity in future work. To support researchers in advancing this future work, we have publicly released the anonymized data and the platform at \textbf{\url{https://social-dynamics.net/fears-and-hopes}}.
\section{Related Work}
\label{sec:related-work}

We surveyed various lines of research that our work draws upon, and grouped them into two main areas for collecting views on AI (\S\ref{subsec:collecting-views}), and analyzing them (\S\ref{subsec:analyzing-views}). 

\subsection{Collecting Views on AI}
\label{subsec:collecting-views}

Historical power imbalances have resulted in some groups gaining greater access to the benefits of AI, while others are more likely to face its risks~\cite{birhane2022power, sloane2022participation, dennler2023bound}. To ensure that AI development and regulation are equitable and inclusive, it is crucial to gather the perspectives of all these groups. Researchers achieve this through both direct and indirect methods \cite{Floridi2018, nader2024public}. Direct methods explicitly ask for opinions through surveys, deliberative processes like community juries and assemblies, in-depth interviews, co-design studies and online narrative studies~\cite{methodsForCollectingAIOpinions2023, kapania2022because, nader2024public, emami2023understanding, laffer2022using}. Indirect methods capture views in their natural settings or as they emerge organically in discussions or behaviors \cite{hohendanner2023exploring, Miyazaki2024}.

The most common direct method for gathering views on AI is through surveys, which provide valuable population-level insights but often limit participant agency. These surveys range from large-scale studies across multiple countries~\cite{scantamburlo2023artificial, kelley2021exciting, googleAIsurvey} to more focused ones targeting representative groups within a single country~\cite{cave2019scary} or specific subgroups, such as AI practitioners, employees and students~\cite{sartori2023minding}. They may employ customized questionnaires or standardized tools like the AI Attitude Scale (AIAS), the Perceptions on AI by the Citizens of Europe (PAICE) questionnaire~\cite{scantamburlo2023artificial}, or the Technology Acceptance Model (TAM) questionnaires~\cite{AIAcceptanceFactors2023}. The focus of these surveys also varies, with some assessing general attitudes towards AI, while others exploring perceptions of specific AI technologies, such as remote biometric identification ~\cite{kieslich2024regulating} or facial recognition \cite{frtCrossCountryPerceptions_2023}. However, surveys alone may not fully capture the depth of people's views or the reasoning behind them. To address this, they are often combined with additional direct methods such as interviews or focus groups to provide a richer, more nuanced understanding. \citet{adus2023exploring} conducted virtual focus groups to understand of how patients can and should be meaningfully engaged within the field of AI development in healthcare. Kapania's study~\cite{kapania2022because} on AI perceptions in India supplemented survey data with in-depth interviews of adult Internet users. Similarly, \citet{nader2024public} combined an online survey with focus groups of entertainment media creators to explore how media shapes public beliefs about AI. However, interviews and focus groups often structure participation in ways that limit spontaneous deliberation and collective sense-making. To foster more dynamic and inclusive discussions, alternative methods such as deliberative polling were explored. For example, the Polis system \cite{polis_2018} helped identify distinct opinion groups by enabling participants to vote on or contribute preferences for AI chatbot behavior \cite{constitutionalAI_2024}. Similarly, the Stanford Online Deliberation Platform \cite{stanfordDeliberationPlatform_2021} facilitated Taiwan's national deliberation by allowing the public to discuss policy options for addressing deepfakes and forged media \cite{taiwanDeliberation_2024}.

Indirect methods for gathering views on AI often involve analyzing secondary sources \cite{Ouchchy2020} or engaging participants in imaginative and interactive scenarios. For instance,~\citet{cave2019hopes} analyzed over 300 news articles, literary works, and films to identify eight fundamental hopes and fears about AI in Western societies. \citet{awad2020approach} had participants imagine solutions to AI-related dilemmas using "Dilemma Vignettes," while ~\citet{robb2020robots} paired a public exhibition with a quiz to explore public perceptions of robots. Another approach is seen in the Collective Constitutional AI (CCAI) project~\cite{huang2024collective}, which gathers preferences for large language model behavior by allowing participants to contribute their preferences and vote for the preferences of others using options like ``Agree'', or ``Disagree''. Some studies combine direct and indirect methods; for example, ~\citet{hohendanner2023exploring} surveyed computer science students and also conducted a speculative design workshop with designers to explore AI perceptions from multiple angles.

\subsection{Analyzing Views on AI}
\label{subsec:analyzing-views}

Once views on AI are collected, they can be analyzed and compared to understand the different perspectives and priorities held by various subgroups. This diversity of views on AI is explored in \citet{jakesch2022different}, where the authors analyze and compare how the U.S. population, crowdworkers, and AI practitioners perceive and prioritize aspects of Responsible AI (RAI). Similarly, \citet{kapania2022because} collected the views of Indian citizens on AI, highlighting how their optimism contrasts with Western skepticism. Recognizing this range of opinions, several studies propose methods to incorporate the perspectives of various stakeholders, including the public, into the AI design process. For example, \citet{lee2019webuildai} introduces WeBuildAI, a participatory design framework that balances differing views by allowing participants to vote on others' perspectives. Additionally, STELA provides a methodology for community-centered norm elicitation to ensure AI value alignment \cite{bergman2024stela}, while the AI Failure Cards help foster a deeper understanding of AI failures and gather preferred mitigation strategies~\cite{tang2024ai}. \citet{situateAIGuidebook_2024} developed a toolkit to guide early-stage discussions on whether, and under what conditions, the development or deployment of an AI system should proceed in the public sector. 

However, incorporating this range of views presents challenges, as noted by \citet{varanasi2023currently}, including difficulties in engaging with and prioritizing unfamiliar perspectives. Furthermore, some cases can result in what is known as ``the tyranny of the majority'', where the interests of the majority are enforced above those of the minority, potentially leading to the systematic disadvantaging of minority groups~\cite{awad2020approach}. 

\smallskip
\noindent\textbf{Research Gap.} Previous works have concentrated on collecting views from either the public or AI practitioners. However, it is typically individuals in positions of influence, such as industry leaders and academics, who drive decisions regarding AI. This leaves an important gap in understanding whether the views of the public are aligned with those of the AI influencers who shape the direction of AI development and policy.

\section{Researcher Positionality Statement}
Before outlining our methodology, we first position ourselves in relation to the research we present in this work. The team consists of individuals with expertise in Data Visualization, Computer Science, and AI. Our team consists of three men and one woman, bringing together diverse experiences from both industrial research labs and academic institutions. We have cultural and professional backgrounds spanning Europe, North America, and South America. We also represent a range of religious affiliations. We acknowledge that our positionality may influence various aspects of our research, including but not limited to our user-study methods, design decisions, aesthetic choices for the platform, language selection, and topics emphasized in qualitative analyses. We recognize the importance of including a broader range of voices from academia and the public to bring unique perspectives to the work we develop.

\section{Methods for Collecting and Comparing AI Views from AI Influencers and the Public}
\label{sec:method}
\begin{figure*}[t!]
  \includegraphics[width=\textwidth]{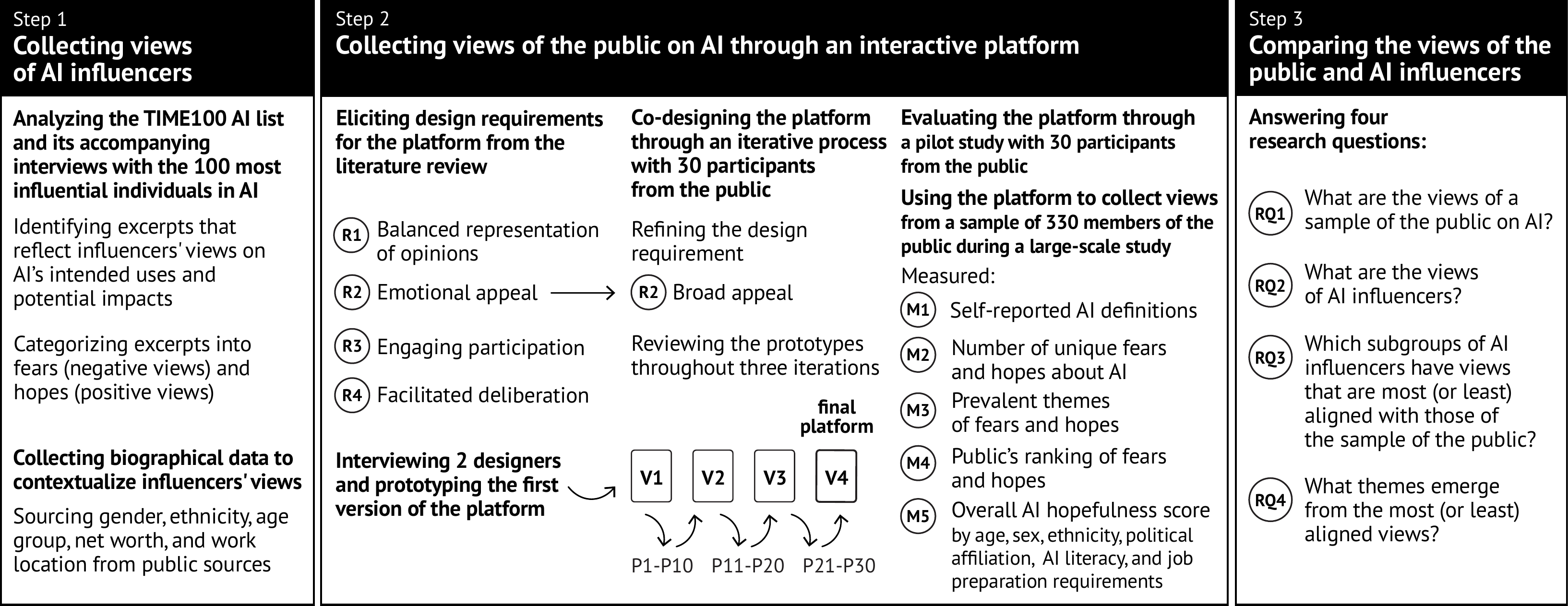}
  \caption{\textbf{Overview of our three-step method for comparing the public's views and AI influencers' views on AI using an interactive platform}. In the first step, we collected views from AI influencers--Time100 nominees \cite{ai100Time}--by analyzing their interviews, focusing on key excerpts about AI's uses and impacts, and categorizing them into hopes and fears. In the second step, we built a platform to collect views from the broader public, starting with a literature review and interviews with two designers to elicit four design requirements and create the initial prototype (V1). Next, we engaged individuals from the public to iteratively update one design requirement and co-design three iterations of the platform (V2-V4). Next, we conducted a pilot study to evaluate the platform and identify improvements for the large-scale study. Finally, we used the platform to gather views from 330 U.S. individuals, representative of the population by age, sex, ethnicity, and political affiliation, capturing their fears and hopes about AI. In the third step, we compared the collected views.}
  \Description{Flowchart with three boxes placed side by side from left to right, each representing a distinct step in the methodology for comparing the views of AI influencers and the public on artificial intelligence.
  Step 1: Collecting Views of AI Influencers. This box describes the process of analyzing the Time100 AI list of influential figures in AI and their related interviews. The process began with identifying excerpts from interviews that reflect hopes (positive views) and fears (negative views) about AI's intended uses and potential impacts. Next, these excerpts were categorized into themes of hopes and fears. To provide contextual background, biographical data for each AI influencer--including gender, ethnicity, age group, net worth, and work location--was then sourced from public information.
  Step 2: Collecting Views of the Public on AI. This box describes the process of gathering public opinions through an interactive platform, developed via an iterative co-design process with 30 participants from the public. The process began with identifying four literature-driven design requirements: broad appeal, balanced representation of opinions, emotional engagement, and facilitated deliberation. Based on these requirements, the platform's design was refined over three iterations, incorporating feedback from 30 participants to enhance its usability and effectiveness for data collection. After finalizing the design, a large-scale study involving 330 participants from the public was conducted to collect their views on AI.
  Step 3: Addressing Four Research Questions. This box outlines the main research questions explored in this study: What are the views of AI influencers? What are the views of the public on AI? Which subgroups of AI influencers have views most or least similar to those of the public?; and what themes emerge from the most or least similar views? 
  To address these questions, the authors defined and collected the following metrics: self-reported AI definitions, number of unique fears and hopes about AI, prevalent themes in fears and hopes, public's ranking of fears and hopes, and overall AI hopefulness score by age, sex, ethnicity, political affiliation, AI literacy, and job preparation requirements.}
  \label{fig:methodology}
\end{figure*}

Next, we describe our three-step methodology (Figure \ref{fig:methodology}) for: collecting views from AI influencers (\S\ref{subsec:method1}); gathering public's views on AI using an interactive platform (\S\ref{subsec:method2}); and analyzing both public's and AI influencers' views (\S\ref{subsec:method3}).

\subsection{Methods for Collecting Views from the AI Influencers}
\label{subsec:method1}

We defined three main criteria for selecting a list of influencers to analyze views on AI: (1) demographic diversity, (2) sectoral coverage, and (3) professional influence. To evaluate demographic diversity, we measured the percentage of women, men, and nonbinary individuals; the percentage of white individuals and people of color; and the percentage of participants within each of six generational cohorts. For sectoral coverage, we examined the distribution of roles across academia, government, industry, and other sectors, and the percentage of participants professionally active in different geographic locations (i.e., based in Silicon Valley \emph{vs.} other locations). Finally, for professional influence, we identified the percentage of billionaires and millionaires as well as the percentage of individuals holding C-suite positions.

We sourced three candidate lists for comparison: the list of members from the UNESCO High-level Advisory Body on Artificial Intelligence~\cite{unesco_list}, which includes 39 individuals; the list of members from the OECD Working Party on Artificial Intelligence Governance~\cite{oecd_list}, which consists of 14 individuals; and the Time100 AI list by Time magazine~\cite{ai100Time}, which features 100 of the most influential individuals in AI. While these lists differ in size and focus, there is some overlap: five individuals appear on both the Time100 AI list and the UNESCO list, and one individual appears on both the Time100 AI list and the OECD list. These lists performed as follows across our criteria:

\begin{itemize}
    \item \textbf{Criterion 1: Demographic diversity}: The OECD list had the highest proportion of women (57\%) but no representation of nonbinary individuals, with the remaining 43\% being men. The UN list followed with 51\% women and 49\% men, also lacking nonbinary inclusion. The Time list balanced representation across genders with 39\% women, 59\% men, and 2\% nonbinary individuals. In terms of ethnicity, the OECD list split evenly between white participants (50\%) and people of color (50\%) but excluded Black or African American individuals. The UN list had the highest proportion of people of color (69\%) but included fewer white individuals (31\%). The Time list struck a middle ground, with 52\% white  and 48\% people of color. Time100 AI provided the broadest age representation, spanning from younger individuals aged 18-24 (1\%) to older individuals aged 65+ (6\%), with the largest group being 35-44 (40\%). In contrast, both the OECD and UN lists were heavily skewed toward professionals aged 35-44, accounting for 57\% and 41\% of their participants, respectively.
    \item \textbf{Criterion 2: Sectoral coverage}: The OECD list was dominated by government representatives (71\%) with limited roles in industry (22\%). The UN list leaned heavily toward academia (41\%) and government (31\%) but underrepresented industry leaders (7\%). The Time list features 31\% of influencers from the industry, and significant representation across academia (19\%) and the arts (11\%). For geographic distribution, 100\% of OECD influencers, 95\% of UN influencers, and 69\% of Time100 AI influencers were active in other locations outside Silicon Valley.
    \item \textbf{Criterion 3: Professional influence}: The Time100 AI list had the highest proportion of financially influential figures, with 11\% billionaires and 31\% millionaires. The UN list included 3\% billionaires and 8\% millionaires, while no data was available for the OECD list. Additionally, 33\% of the Time100 AI comprised C-suite executives, significantly more than the UN list (13\%) and the OECD list (which had none).
\end{itemize}

Based on these comparisons, we selected the Time100 AI list for our analysis. It offered the most balanced demographic diversity, broadest sectoral coverage, and strongest representation of professionally influential individuals. To verify these statistics, we contacted the editorial team of Time magazine who provided detailed information on their selection and interview process. The list was developed through an editorial process including more than five dozen reporters and editors specializing in AI and business reporting, with input from field experts. The process, spanning 4-5 months, began with research and nominations in April 2023 and culminated in the list's release in August 2023. Influencers were nominated based on notable achievements such as publishing influential research, developing impactful products, or driving societal change through regulatory actions. The team evaluated a few hundred candidates and narrowed them down to the final 100 through weekly reviews, ensuring thematic diversity between sectors, roles, and regions for inclusion. These nominees were interviewed via video or phone, with discussions tailored to their individual achievements and broader impact. The resulting interviews were published on a dedicated webpage, with quotes clearly labeled as direct or paraphrased. In cases where interviews incorporated statements from publicly available materials such as YouTube videos, these sources were clearly attributed to ensure transparency.

The Time100 AI list features 43 CEOs, founders, and co-founders, and 41 women and nonbinary individuals, with ages ranging from 18 to 76. The nominees are grouped into four categories: leaders (e.g., founders), innovators (e.g., artists), shapers (e.g., political advisors), and scholars (e.g., researchers). The list includes short bios of the nominees and long interviews that capture their perspectives on AI's current state and future development. To collect their views, three authors read the interviews and independently marked excerpts that focused on AI's intended uses and potential impacts. They then jointly categorized these excerpts into hopes and fears and assessed the overall sentiment of each article. 

To assess the sentiment, they searched for direct expressions of emotions (e.g., fear, anticipation) as per Plutchik's wheel of emotions~\cite{plutchik2013theories} and previous works exploring human emotions at scale \cite{WeFeelFine}. Plutchik's wheel of emotions was favored over alternatives (e.g., Ekman's six emotion model~\cite{ekman1992argument}, or the positive-negative activation (PANAS) model of emotion by Watson and Tellegen~\cite{watson1985toward}) because it provides a wider range of emotions and highlights relationships between them, offering flexibility in analyzing complex emotional expressions.
However, we acknowledge it fails to account for contextual factors like personal history, social norms, or situational influences as emotional expression and interpretation vary across cultures. If direct expressions were not present, the authors inferred the sentiment based on contextual clues and overall tone. In cases where the emotional content was unclear or there were disagreements, the authors discussed these instances and collaboratively determined the most dominant emotion. To enrich our dataset, we extended Time's magazine bios by including additional biographic data such as gender, ethnicity, age group, net worth, and current working location (e.g., Silicon Valley or other locations). This biographic data was manually extracted from public company portals.

\subsection{Methods for Collecting AI Views from the Public}
\label{subsec:method2}

\subsubsection{Designing an Interactive Platform to Collect Views from the Public on AI}
The platform served as a data collection tool to answer our four RQs. To design it, we followed three steps (Figure \ref{fig:methodology}). These are: \emph{1)} a literature review; \emph{2)} interviews with two designers employed at a large tech company and in academia; and \emph{3)} and an iterative co-design process with 30 participants (representative of the public) recruited from Prolific \cite{prolific}. 

\mbox{}
\newline
\noindent\textbf{Step 1: Literature Review.} To gather an initial set of requirements for the visualization, we conducted a targeted review of key papers focused on surveying public opinion on AI. This review was not intended to exhaustively cover all relevant literature but served as a starting point to inform our design, following similar co-designing practices \cite{raiGuidelines2024, impactAssessmentTemplate2024, socieltalImpactTemplate2024}. Specifically, we formulated two questions to guide our literature search, that is: \emph{What challenges do participants face when providing views about AI?}; and \emph{What features do participants prefer in tools for collecting and visualizing these views?}

\aptLtoX[graphic=no,type=html]{\begin{titled-frame}{\textcolor{white}{\textit{Search Query}}}
 \noindent\textbf{Subject:} (ARTIFICIAL INTELLIGENCE) \\
      \noindent\textbf{What:} (PUBLIC OPINION OR PUBLIC PERCEPTION OR PUBLIC ATTITUDE OR PUBLIC VIEW OR PUBLIC ACCEPTANCE) \\
      \noindent\textbf{How:} (SURVEY OR TOOL OR VISUALIZATION)
\end{titled-frame}}{\vspace{0.65cm}
\begin{center}
  \setlength{\fboxsep}{12pt}
  \setlength{\fboxrule}{0.65pt}
  \fcolorbox{black}{white}{
    \begin{minipage}{22.7em}
    \vspace{0.2em}
      \noindent\textbf{Subject:} (ARTIFICIAL INTELLIGENCE) \\
      \noindent\textbf{What:} (PUBLIC OPINION OR PUBLIC PERCEPTION OR PUBLIC ATTITUDE OR PUBLIC VIEW OR PUBLIC ACCEPTANCE) \\
      \noindent\textbf{How:} (SURVEY OR TOOL OR VISUALIZATION)
    \end{minipage}
  }
  
\makebox[0pt][l]{
  \hspace*{-12.25em}
  \raisebox{8.95em}[0pt][0pt]{
    \setlength{\fboxsep}{5pt}
    \fcolorbox{black}{black}{
      \parbox[c][0.8em][c]{6em}{
        \centering
        \textcolor{white}{\textit{Search Query}}
      }
    }
  }
}
\end{center}}

\smallskip
We searched for papers in the ACM Digital Library \cite{acm} using a three-components query, involving: \emph{Subject}, \emph{What}, and \emph{How} (as shown in the box above). This search identified a total of 120 papers.
We then screened these papers using inclusion criteria. Papers were included if they: \emph{1)} discussed public opinion, perception, views, or attitudes towards AI; \emph{2)} presented tools for gathering and visualizing public views; and \emph{3)} included user requirements or recommendations for future tools. This procedure left us with 8 papers \cite{cave2019scary, zhang2020us, robb2020robots, kelley2021exciting, hohendanner2023exploring, sartori2023minding, doom2023, kieslich2024regulating}.

We then conducted a full-text analysis of these papers to document challenges and preferences for the platform. We highlighted key excerpts related to challenges, opportunities, and user requirements, and grouped these excerpts into common themes such as public engagement, usability, and accuracy of view representation. Each theme included excerpts from at least two papers, signifying that data saturation was reached~\cite{guest2006many}. From these themes, we derived four design requirements for the platform:
\begin{enumerate}[left=0pt]
    \item [R1] \emph{Balanced representation of views}: Ensure that both positive and negative views are presented in the platform.
    \item [R2] \emph{Emotional appeal}: Use visual elements that increase public's sense of contribution and agency.
    \item [R3] \emph{Engaging participation}: Make the interaction with the platform interactive and engaging.
    \item [R4] \emph{Facilitated deliberation}: Enable the public to form well-rounded opinions about AI.
\end{enumerate}
\smallskip

\noindent\textbf{Step 2: Interviews with Designers.} 
Having the requirements at hand, we conducted 30-minute semi-structured interviews with two designers---one male and one female, aged 24 and 32, with 4 and 10 years of experience in data visualization, respectively, employed in academia and at a large tech company---to explore potential ways to visualize these requirements and prototype the first version of the platform (Appendix \ref{sec:interviews_designers}). To recruit these designers, we used a responsible AI email list at a large technology company and adopted a purposive sampling approach \cite{purposiveSampling2020}. This approach allowed us to intentionally target designers possessing the qualifications necessary for the study, particularly those with expertise in responsible AI, HCI, and data visualization, who: (1) have experience in analyzing large-scale public opinion data, including tools for collecting, interpreting, and visualizing public sentiment, and (2) publish their findings in peer-reviewed venues such as academic conferences. Out of five designers who expressed interest, two were available to participate within the project's timeline.
\smallskip

\noindent\textbf{Step 3: Co-design Process.} After designing the first version of the platform, we conducted three co-design sessions with 30 participants from the public who were recruited from Prolific \cite{prolific}. These participants were selected to reflect the U.S. population demographics in terms of gender, \cite{census_ethnicity_2020, census_age_gender_2022} (14 males and 16 females), with ages ranging from 18 to 64 years, and included 19 White, 3 Black, 1 Asian, and 8 individuals of Mixed or Other ethnicities. They were also digitally literate, exposed to visual media, and skilled in communication for providing feedback. Participants were asked to carefully review the prototype, consider how they would interact with the platform if fully functional, and then provide feedback on what worked well, what was confusing or needed improvement, and whether the design would encourage them to share their views on AI (Appendix \ref{sec:codesign_questions}).

The study was approximately 30-45 minutes long and participants were paid on average about \$12 (USD) per hour. We then conducted inductive thematic analysis \cite{miles1994qualitative}, examining the recommendations for the next version of the platform. In total, we made three iterations, expanding our second design requirement R2 from \emph{Emotional appeal} to \emph{Broad appeal} to ensure the platform resonated with people of varying AI knowledge levels. We iteratively refined the design based on participants' recommendations (details in Appendix \ref{sec:design_iterations}), leading to its final version.
\smallskip

\begin{figure*}[!tb]
    \centering
    \includegraphics[width=0.89\textwidth]{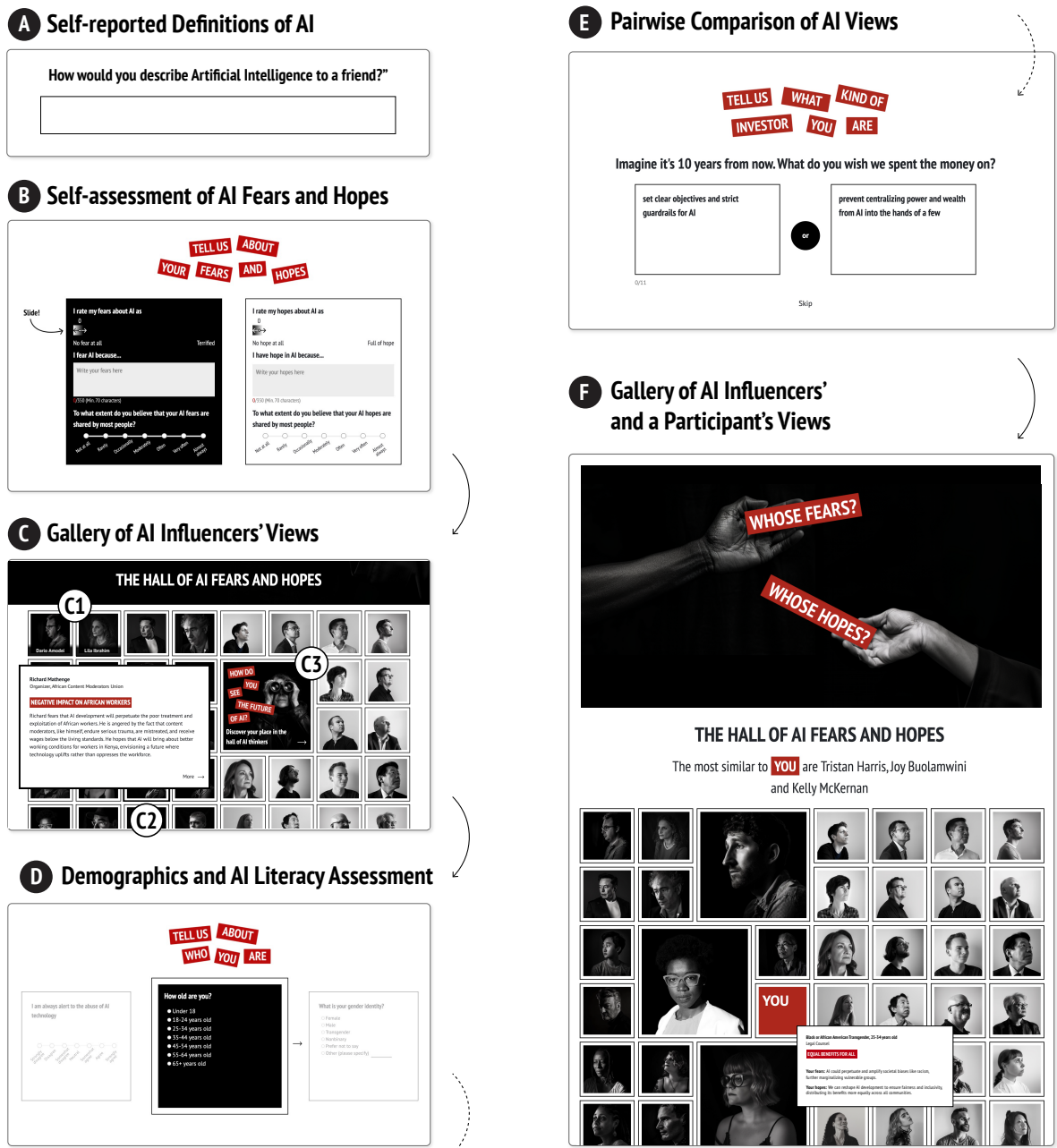}
    \caption{\textbf{The ``The Hall of AI Fears and Hopes'' is an interactive platform consisting of six sections for collecting public views on AI and presenting the views of AI influencers.} In the first section (A), participants share their definitions of AI. In the second section (B), they rate their fears and hopes, describe their specific fears and hopes, and indicate how widely they believe these views are shared. In the third section (C), participants can view portraits of AI influencers, organized by their more negative (left) and more positive (right) attitudes toward AI. By hovering (C1) and clicking on a portrait (C2), more information about the influencers' fears and hopes appears in a pop-up box. Participants proceed to the next section (D) by clicking on (C3), where they provide data on gender, ethnicity, age, education level, occupation, and AI literacy. In the fifth section (E), participants imagine a future 10 years ahead and vote on how to allocate money based on 10 randomly selected pairs of AI influencers' fears and hopes. In the final, sixth screen (F), participants see their symbolic portrait displayed alongside AI influencers, positioned based on their fear and hope ratings, as well as the similarity of their fears, hopes, and demographics (age, gender, ethnicity) to those of the influencers.}
    \Description{The Hall of AI Fears and Hopes is an interactive platform designed to collect public views on AI and present views from AI influencers. It consists of six vertically arranged sections, labeled from A to F. Participants can move between sections using navigation buttons, with each section building on the previous one.
    In section A, participants share their definitions of AI. In section B, they rate their fears and hopes on a scale ranging from 1 (no fear or hope at all) to 10 (terrified or full of hope), describe their specific fears and hopes, and indicate how widely they believe these fears or hopes are shared, using a scale from 1 (not at all shared) to 7 (almost always shared). In section C, participants see portraits of AI influencers, arranged based on their attitudes toward AI, from more negative on the left to more positive on the right. Hovering over a portrait reveals the influencer's name, and clicking opens a pop-up with their fears and hopes. Participants proceed to the next section by clicking on the largest portrait with a provocative message: Discover your place in the Hall of AI Thinkers. In section D, participants provide their demographic data, including gender, ethnicity, age, education, occupation, and AI literacy. In section E, participants are asked to imagine the world 10 years from now and vote on how to allocate money based on 10 randomly selected pairs of AI influencers' fears and hopes. In the final section, F, participants are shown their symbolic portrait alongside the portraits of AI influencers. The position of the participant's portrait is based on their fear and hope ratings, as well as the similarity of their fears, hopes, and demographics (age, gender, ethnicity) to those of the influencers.}
    \label{fig:final-viz}
\end{figure*}

\noindent\textbf{Final Visualization Platform.} The platform consists of six sections and balances between collecting the public's views on AI and exposing them to the views held by AI influencers\footnote{``The Hall of AI Fears and Hopes'' is available at \textbf{\url{https://social-dynamics.net/fears-and-hopes}}} (Figure \ref{fig:final-viz}).

In the first section of the platform (\emph{Self-reported Definitions of AI}), we invited participants to share their interpretation of AI by asking ``How would you describe Artificial Intelligence to a friend?'' \cite{cave2019scary}. In the second section of the platform (\emph{Self-assessment of AI Fears and Hopes}), participants are given two cards: one black card on the left for AI fears, and one white card on the right for AI hopes (Figure \ref{fig:final-viz}A). On each card, they rate their level of fear or hope on a scale from 0 (``No fear / No hope'') to 10 (``Terrified / Full of hope''). They also write down their specific fears and hopes and indicate how strongly they believe others share these views. In the third section (\emph{Gallery of AI Influencers' Views}), participants are presented with black-and-white portraits of AI influencers alongside their views (Figure \ref{fig:final-viz}B; see \S\ref{subsec:method1} for how these views were collected from the interviews). The portraits are organized into two columns: on the left, influencers with more fearful views appear on a black background, while on the right, influencers with more hopeful views appear on a white background. When participants hover over a portrait, the influencer's name and surname appear (Figure \ref{fig:final-viz}B1). When they click on a portrait, the influencer's organizational role is displayed, along with their hopes and fears, and a link to the full interview (Figure \ref{fig:final-viz}B2). One portrait in the gallery stands out—it is enlarged with red labels, inviting participants to click and ``Discover your place in the hall of AI thinkers'' (Figure \ref{fig:final-viz}B3). Once clicked, participants are taken to the fourth section (\emph{Demographics and AI Literacy Assessment}) as shown in Figure \ref{fig:final-viz}C. In this section, participants provide information on their gender, ethnicity, age, education, occupation, job training and education requirements, and AI literacy. The job training and education requirements are based on O*NET's job zones, which classify occupations by the level of education, training, and experience required, ranging from zone 1 (minimal education and training, e.g., dishwashers) to zone 5 (extensive education training, e.g., surgeons) \cite{jobZones_ONET}. For clarity, we rephrased O*NET's original ``job zones'' as ``job training and education requirements''. AI literacy is measured using the AI Literacy Scale (AILS) \cite{AILiteracyScale2023}. This information, along with the fears and hopes self-assessment from the first section, will be used to match and present the participants' views alongside those of AI influencers in the final, sixth section of the visualization. In the fifth section (\emph{Pairwise Comparison of AI Views}), participants are asked to imagine a future 10 years from now and answer ``What do you wish we spent the money on?'' (Figure \ref{fig:final-viz}D). 
They are then shown 10 randomly selected pairs from a set of 20 hopes and 20 fears expressed by AI influencers. Each pair of items may consist of two hopes, two fears, or one hope and one fear, enabling many contrasting comparisons.\\Participants have the option to skip any pair, if they do not wish to choose. To avoid any predictable patterns in the comparison process, each pair was presented in a randomized order. Additionally, the order of items within each pair (e.g., hope on the left, fear on the right) was randomized in each round to control for potential biases related to item positioning. We chose this pairwise comparison method over alternatives like Likert scales for two reasons \cite{salesses2013collaborative}. First, it reduces participants' cognitive load, as comparing two items at a time is simpler than rating many at once. Second, it allows us to generate a ranked list of hopes and fears that reflects the preferences of the members of the public. After completing the pairwise comparison, participants move to the final, sixth section (\emph{Gallery of AI Influencers' and a Participant's Views}) as shown in Figure \ref{fig:final-viz}E. This section displays both AI influencers' portraits and the participant's portrait. The participant's portrait is positioned based on their fear and hope ratings from the first section, the similarity of their fears and hopes to those of the influencers, and the alignment of their demographics (i.e., age, gender, ethnicity) from the fourth section with those of the influencers. Horizontally, portraits are placed on a spectrum from fearful to hopeful, while vertically, the placement is determined by the cosine similarity between the participant and their top three most similar influencers, considering both their written responses and demographic alignment. By positioning the participant among their most similar influencers, we apply the concept of homophily—the idea that people tend to connect more with those who are similar to them. Research shows that individuals often look to those who are alike when forming opinions and making decisions \cite{socialComparisons_1954, similarPeople_2017}. Therefore, we used this principle to strengthen the participant's connection to the displayed opinions and encourage a more thoughtful and engaged reflection. Overall, the platform's design emphasizes the fear \emph{vs.} hope framing through a black-and-white color palette, while the intentional use of realistic portraits humanizes the concept of AI by empowering participants to see AI as shaped by human perspectives.

\subsubsection{Evaluating the Platform for Collecting Views from the Public about AI}
\label{subsubsec:viz-better-than-survey}

We conducted a pilot study with 30 new participants, reflecting U.S. demographics in terms of sex and ethnicity to evaluate the platform's design and simulate the upcoming large-scale study \cite{pilotStudy2024, tahaei2024surveys}. This pilot study assessed four design requirements: balanced representation of views (R1), broad appeal (R2), engaging participation (R3), and facilitated deliberation (R4). Participants rated these requirements on a 1-7 Likert scale, with average scores ranging from 4.94 to 5.88, indicating general agreement that the platform effectively met these design goals. Key areas for improvement included navigation and question phrasing. In response, we added arrow annotations, refined question wording, and updated survey settings, including adjusted monetary rewards for older participants, to enhance the platform's usability and response quality. These updates ensure the platform is ready for the large-scale study. Full demographics, methodological details, ratings, and a thematic analysis of participant feedback are provided in Appendix \ref{sec:pilot_evaluation}.

\subsection{Methods for Comparing the Views of the Public versus AI Influencers}
\label{subsec:method3}

We conducted a large-scale user study with 330 participants (representing the public) to collect their views on AI and expose them to the views of AI influencers. Next, we describe our study's design: its setup (\S\ref{sec:setup}); research questions, metrics and analysis (\S\ref{sec:metrics}); and execution (\S\ref{sec:execution}).

\subsubsection{Setup}
\label{sec:setup}
The study consisted of three steps. In the first step, participants received a brief introduction to the study and completed a warm-up task in which they described Artificial Intelligence to a friend. We chose this question based on a study of the public's fears and hopes in the UK \cite{cave2019scary} to ease participants into the survey and smoothly transition to more complex, personal questions about their fears, hopes and AI literacy. In the second step, participants interacted with our platform. We incorporated three attention checks directly into the platform to ensure that participants were engaged. In the third step, participants rated how well the platform met the four design requirements we identified with the members of the public and provided any additional feedback.

\subsubsection{Research Questions, Metrics, and Analysis}
\label{sec:metrics}

Upon that procedure, we answered four research questions.
\smallskip

\noindent{\emph{RQ\textsubscript{1}:} What are the views of a sample of the U.S. public on AI?} \\
\noindent We answered this question both quantitatively and qualitatively. Quantitatively, we defined two metrics: 

\begin{itemize}

\item The number of unique fears and hopes. We computed this in four steps. First, we collected all sentences of hopes and fears provided by participants in the first section of the platform (Figure~\ref{fig:final-viz}A) and generated their embeddings using the sentence transformer model `all-mpnet-base-v2' \cite{sentenceTransformer}. We chose this model for its high performance on single sentences and short paragraphs \cite{song2020mpnet}—the types of texts participants used to formulate their hopes and fears— and for its ability to produce high-quality embeddings. These embeddings were essential for the second step: clustering sentences to identify and remove duplicate fears and hopes. For this, we used HDBSCAN \cite{campello2013density}, a density-based clustering algorithm that does not require specifying the number of clusters, can manage clusters of different sizes and densities, and detects outliers (non-duplicates). We ran the algorithm to generate the initial set of clusters along with an outlier group. Two of the authors manually reviewed all sentences by visualizing them on the Figma platform \cite{figma}, agreeing on where outliers should be placed and finding opportunities to merge clusters. Third, from each cluster, we randomly selected one fear or hope to keep as a representative, since the sentences in each cluster were very similar, and deleted the rest. For example, consider the two sentences: ``AI will take away your own thinking and creativity'' and ``We will lose skills in writing and creativity''. These sentences convey the same fear, so only one would be kept as the representative, while the other would be removed. Finally, we counted the resulting overall number of unique hopes and that of fears. 

\item The hopefulness score for particular subgroups of the public. We analyzed different subgroups of the public (e.g., males \emph{vs.} females; younger participants, defined as those under 38 years old—about half the median life expectancy in the U.S.—\emph{vs.} older participants, defined as those above 38 years old). Each participant $i$ within a subgroup $s$ provided two ratings in the second section of the platform (Figure~\ref{fig:final-viz}B): $fear_{rating}$ and $hope_{rating}$. 
Then, for each participant in the subgroup, we computed an individual hopefulness score by subtracting their self-reported $fear_{rating}$ from their $hope_{rating}$. Finally, the $\text{hopefulness\_score}_{\text{s}}$  was obtained by averaging the individual scores of all participants within the subgroup (Equation~\S\ref{eq:hopefulness_score}), with possible values ranging from $[-10,10]$.

\begin{equation}
\text{hopefulness\_score}_{\text{s}} = \frac{1}{N_s} \sum_{i=1}^{N_s} \left( \text{fear\_rating}_i - \text{hope\_rating}_i \right)
\label{eq:hopefulness_score}
\end{equation} 

where $N_s$ is the number of participants in subgroup $s$; 

$\text{fear\_rating}_i$ and $\text{hope\_rating}_i$ are the fear and hope ratings provided by participant $i$.
\end{itemize}

Qualitatively, we thematically analyzed participants' self-reported definitions of AI, along with their associated fears and hopes. To analyze the definitions, we adopted two complementary approaches. First, we categorized the definitions into clusters \cite{AIRoles_2023}. This involved preprocessing the definitions by removing repetitive phrases such as ``Artificial Intelligence is...'' and generating text embeddings using the `sentence-transformers/all-mpnet-base-v2' model \cite{sentenceTransformer}. K-means clustering was then applied to these embeddings, with the optimal number of clusters ($k = 4$) determined using the elbow method. Second, we examined the linguistic components of the definitions \cite{cave2019scary, kelley2021exciting, AICapabilities2023}. This involved tokenizing the definitions and analyzing the relationships between key linguistic elements such as nouns, adjectives, and verbs. For fears and hopes, we focused on identifying the most prevalent themes, and examining differences among specific subgroups of the public. Prevalent themes were derived through inductive coding, where two authors systematically reviewed unique fears and hopes, grouping them into themes based on established qualitative methodologies~\cite{saldana2015coding, miles1994qualitative, mcdonald2019reliability}. We included themes with input from at least ten participants and reported their occurrences, with participant quotes marked with ``P''. Differences among specific public subgroups were identified by analyzing their hopefulness scores in two ways. First, we examined opposing perspectives within subgroups by dividing them into segments with contrasting average scores (positive \emph{vs.} negative). Second, we analyzed outlier perspectives within high-variance subgroups, focusing on participants whose scores deviated by more than one standard deviation, representing extreme fearfulness or hopefulness. These differences were further explored through inductive coding \cite{saldana2015coding, miles1994qualitative, mcdonald2019reliability}.

\smallskip
\noindent{\emph{RQ\textsubscript{2}:} What are the views of AI influencers?} \\
\noindent In a way similar to \emph{RQ\textsubscript{1}}, we answered \emph{RQ\textsubscript{2}} in a quantitative and qualitative way. Quantitatively, we calculated the number of unique hopes and fears by following a four-step process. First, we collected all fears and hopes sentences from Time magazine~\cite{ai100Time} using qualitative content analysis, a commonly used method in HCI for analyzing interview data \cite{Krippendorff2019, qualitativeInterviews2023}. To do this, two authors independently analyzed the same subset of 25 interviews, identifying excerpts that reflected influencers' fears and hopes. They started by looking for explicit phrases such as ``[Influencer] fears/hopes that...''. If these phrases were not present, they focused on direct quotes from the interviews. When neither explicit phrases nor direct quotes contained clear fears or hopes, they carefully examined the rest of the interview text for any contextual cues that implied fears or hopes. Second, we conducted a collaborative coding session with the entire research team to compare the independently marked excerpts. During this session, we discussed and resolved discrepancies and iteratively refined a shared coding frame to classify direct quotes and contextual cue excerpts as either fears or hopes. This refined coding frame was subsequently applied to the remaining interviews, with the same two authors independently coding another set of 25 interviews. Throughout this process, the authors maintained ongoing dialogue to resolve uncertainties and ensure consistent application of the coding frame.
Third, to improve clarity and comparability between fears and hopes, two authors employed ethical fabrication \cite{fabrication2012}, making minor adjustments to unify the style of excerpts while preserving their original meaning. For example, they began each fear or hope with an active verb to emphasize the intended action or outcome behind the excerpt. To illustrate, consider this excerpt from an interview with a female CEO: she ``hopes its data can facilitate the preservation of [...] forests and other natural environments''. This excerpt was categorized as a ``hope'' and adjusted to read as ``develop AI to help preserve natural environments''.
Finally, as the fourth step, we identified and removed duplicates and counted the resulting overall number of fears and hopes. Since we did not have any ratings from influencers regarding their individual levels of fears and hopes, we could not compute the hopefulness score for them.

Qualitatively, we thematically analyzed the fears and hopes of influencers through an inductive thematic analysis \cite{saldana2015coding, miles1994qualitative, mcdonald2019reliability}. Starting with a subset of the data, two authors independently performed open coding and discussed their findings with the research team to establish a consensus on code format and granularity. Using this coding framework, the same two authors independently coded the remaining data, and resolved major disagreements through discussion. Finally, the entire research team grouped the codes into key themes. We report them alongside their occurrences and illustrative quotes from influencers, which are marked with ``I'' for clarity.

\smallskip 
\noindent{\emph{RQ\textsubscript{3}:} Which subgroups of AI influencers have views that are most (or least) aligned with our sample of the public?} \\
\noindent We answered this question in a quantitative way by defining four metrics:

\begin{itemize}
    \item The public's ranking of fears and hopes of AI influencers. 
    In the fourth section of the platform, all participants were asked to vote on 10 randomly chosen pairs (Figure \ref{fig:final-viz}D), where the pairs could be either two hopes, two fears, or one of each; these pairs were based on a set of 20 representative fears and 20 representative hopes from AI influencers. For each fear or hope, we calculated a $Q$-score ~\cite{salesses2013collaborative}, a measure reflecting how frequently it was selected over other options in pairwise comparisons. This score was determined using the ``win'' and ``loss'' ratios from all participants comparisons, ensuring a fair and balanced evaluation. We then sorted the fears and hopes in descending order by their $Q$-scores, transforming the pairwise comparisons into an ordered list with the most important fears and hopes of the public at the top and the least important at the bottom.

    \item The ranking of fears and hopes of AI influencers by subgroups of the public.
    We analyzed the voting preferences of mutually exclusive public subgroups (i.e., younger \emph{vs.} older participants). To do this, we filtered the votes from all participants to isolate those from each subgroup. Using these subgroup-specific votes, we recalculated the $Q$-score \cite{salesses2013collaborative} for each fear and hope and then sorted them in descending order to create a ranked list that reflects the preferences of each subgroup.

    \item The misalignment score between between subgroups of AI influencers and the public as a whole.
    We analyzed different, mutually exclusive subgroups of AI influencers (i.e., younger influencers \emph{vs.} older influencers, Silicon Valley workers \emph{vs.} those in other locations, white individuals \emph{vs.} people of color). These binarized forms, although problematic because they generalize complex identities into broad categories--thereby overlooking intersectionality and individual nuances--are required for the statistical significance of our statistical analyses and allow us to systematically compare difference in misalignment for subgroup pairs. To compute the misalignment score, we compared the public's ordered list of fears and hopes with an unordered list of fears and hopes from different AI influencer subgroups, and did so using the method described by \citet{hu2008collaborative} (Equation \S\ref{eq:q_score}):
    \begin{equation}
    \overline{\text{misalignment\_score}}_{\text{s}} = \frac{\sum_{i} \text{importance}_{i} \cdot \text{rank}_{i}}{\sum_{i} \text{importance}_{i}}
    \label{eq:q_score}
    \end{equation}

    The $importance_{i}$ of each sentence $i$ (whether it is a fear or a hope) was determined by the number of times that this fear or hope was mentioned by AI influencers from a subgroup $s$. The $rank_{i}$ represented the percentile position of a sentence ${i}$ on the public's ordered list, with 0\% indicating the top and 100\% the bottom. Lower misalignment score values ($\overline{\text{misalignment\_score}}_{\text{s}}$) indicated that the public ranks sentences considered important by AI influencers closer to the top of their ranked list. In this context, lower values were preferable: a score of 0 signified complete alignment, 0.5 signified no alignment (a result that is essentially random), and 1 signified complete misalignment. When calculating the misalignment score, we accounted for the different number of sentences in the unordered lists of fears and hopes, which varied depending on the subgroup of AI influencers being analyzed (Appendix \ref{sec:misalignment-score}, Figure \ref{fig:misalignment-distribution}). 
    
    \item The misalignment score between subgroups of AI influencers and the matching subgroups of the public. We analyzed how AI influencer subgroups aligned with their public counterparts (i.e., younger influencers with younger participants). To do this, we calculated the misalignment score for each subgroup pair by comparing the unordered list of fears and hopes from the AI influencer subgroup with the ordered list of fears and hopes from the matching public subgroup.
\end{itemize}

\smallskip

\noindent{\emph{RQ\textsubscript{4}:} What themes emerge from the most (or least) aligned views?} \\
\noindent We answered this question in a qualitative way by conducting a thematic analysis of the unique hopes and fears expressed by subgroups of AI influencers whose views were either most aligned or least aligned with those of the public. To do this, two authors first analyzed the misalignment scores between AI influencer subgroups and their corresponding public subgroups, calculated in the previous step (RQ3). This helped them identify which public subgroups aligned most and least with influencers. Next, they examined the hopes and fears in these aligned and misaligned subgroups using an inductive thematic analysis~\cite{saldana2015coding, miles1994qualitative, mcdonald2019reliability}. Each theme was supported by quotes from at least five influencers (marked with ``I'').

\subsubsection{Execution}
\label{sec:execution}

We used Prolific's built-in screeners to control for the participants' geographic location, age, sex, ethnicity, and political affiliation. We used stratified random sampling to represent U.S. census demographics in terms of age (12\% in range 18-24 years, 17\% in range 25-34 years, 17\% in range 35-45 years, 15\% in range 45-54 years, 17\% in range 55-64 years, 13\% in range 65-74 years, and 9\% over 75 years), sex (50\% female, 50\% male), ethnicity (63\% White, 11\% Black, 11\% Mixed, 7\% Asian, 8\% Other), and political affiliation (30\% Republicans, 30\% Democrats, 40\% Independent). We limited our participant pool to individuals residing in the U.S. All participants were paid on average about \$12 (USD) per hour. 

We developed a web-based survey and administered it on Prolific (Appendix \ref{sec:prolific-setup}, Figure \ref{fig:prolific-survey}). The survey comprised four pages, each corresponding to a setup step plus a final confirmation screen. To ensure response quality, we implemented four real-time content validation measures, following best practices for conducting crowdsourcing studies~\cite{checklistCrowdsourcingBiases2021, tahaei2024surveys, pilotStudy2024}. First, we set word ranges of 70-350 characters for open-ended questions not only to ensure participants provided concise, relevant answers but also to prevent survey fatigue. Second, we disabled the ability to paste content from external sources or edit previous responses, ensuring that all answers were original and thoughtfully composed. Third, we tracked clicks on platform's visual elements (e.g., portraits of AI influencers) and monitored time spent on each survey step. 

Fourth, we incorporated three attention checks throughout the survey to identify low-quality responses, in line with Prolific's attention and comprehension check policy \cite{prolific_checks}. In the \emph{Demographics and AI Literacy Assessment} (Figure \ref{fig:final-viz}C), participants encountered two instructional manipulation checks. In the first, they were explicitly instructed to \emph{``Select ``Blue'' when asked for your favorite color''}. In the second, they were told to \emph{``Strongly disagree or disagree with the idea that AI technologies are mainly developed by little squirrels''}. In the \emph{``Pairwise Comparison of AI Views''} (Figure \ref{fig:final-viz}D), participants were asked to vote on a pair of statements, one of which included a nonsensical fear: \emph{`[I wish I had spent the money to] prevent AI from turning all humans into pineapples overnight''}. This statement should not be preferred over the second option, which was drawn from our dataset of the fears and hopes of AI influencers. To be included in the final analysis, participants had to pass all three attention checks and read the fears and hopes of at least five AI influencers.
\section{Comparing Public and Influencer Views on AI}
\label{sec:results}

\subsection*{\emph{RQ\textsubscript{1}:} \textbf{What are the views of our sample of the U.S. public on AI?}}
\label{sec:hopes-fears-broad-public}

\begin{figure*}[t!]
    \centering
    \includegraphics[width=\textwidth]{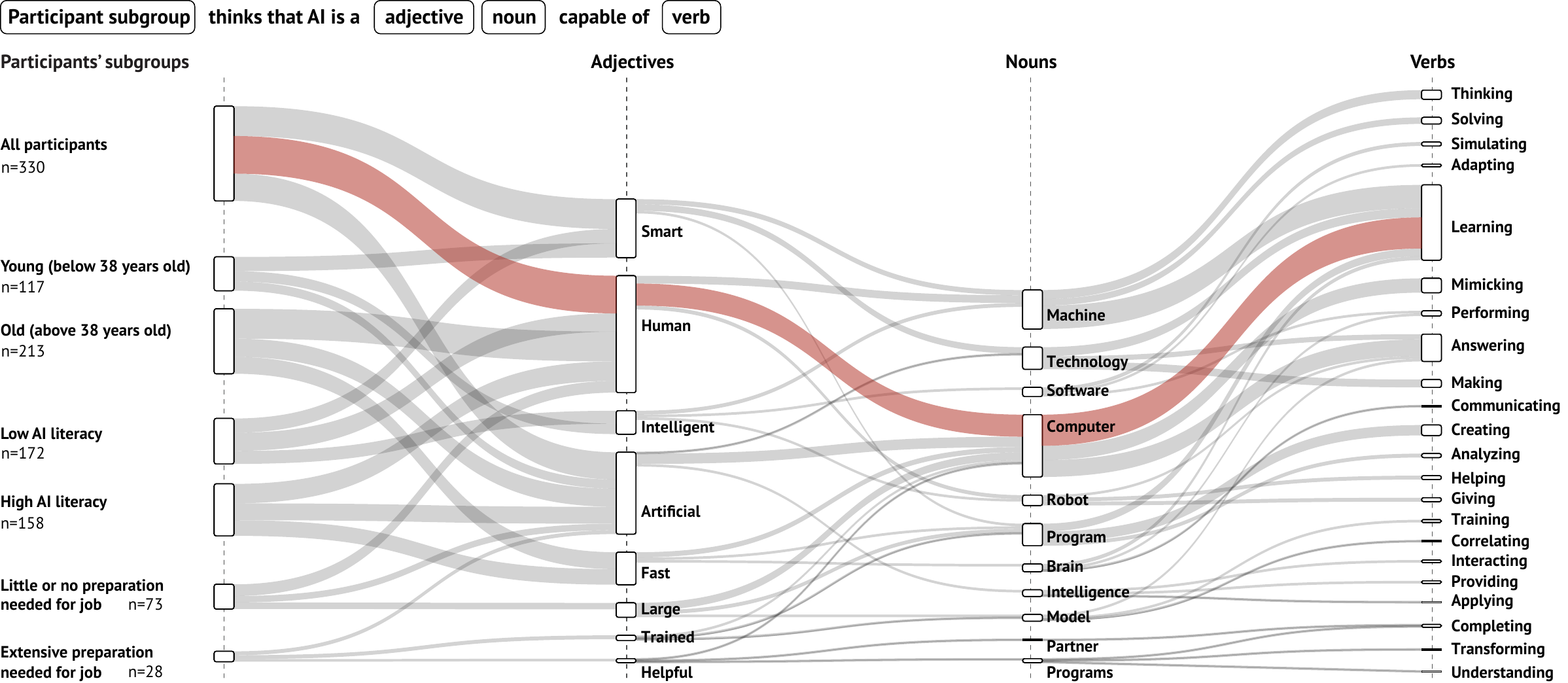}
    \caption{\textbf{Self-reported AI definitions across subgroups in our sample of the U.S. public.} The chart reconstructs these definitions from left to right, highlighting popular adjectives (``human'', ``artificial'', ``smart''), nouns (``computer'', ``machine'', ``technology'', ``program''), and verbs (``learning'', ``answering'', ``creating''). The most common definition across participants is: ``AI is a human-like computer capable of learning'' (path marked in red).}
    \Description{This Sankey diagram illustrates self-reported AI definitions among different participant subgroups from a sample of 330 individuals from the U.S. public. The chart starts on the left side by connecting participant subgroups categorized by age (young below 38 years vs. old above 38 years), AI literacy levels (low vs. high), and job preparation requirements (little or no preparation vs. extensive preparation needed). Then, moving to the right, it decomposes these definitions into horizontal flows of adjectives, nouns, and verbs that reflect how participants describe AI.
    The thickness of each flow represents the frequency of associations, highlighting how groups with differing ages or AI literacy levels perceive AI capabilities differently, using varied descriptive language. The most frequently used adjectives are ``human'', ``artificial'', and ``smart''. The most frequent nouns are ``computer'', ``machine'', ``technology'', and ``program''. The most frequent verbs are ``learning'', ``answering'', and ``creating''. The most frequent definition across participants is: ``AI is a human-like computer capable of learning''.}
    \label{fig:definitions} 
\end{figure*}

\noindent\textbf{Self-Reported Definitions of AI.}  We derived four clusters of AI definitions, ranked from most to least common. The largest cluster, predictive system ($n = 132$), describes AI as an analytical tool capable of processing vast amounts of information to identify patterns and generate predictions. The second most common cluster, human-like companion ($n = 81$), defines AI as a system designed to mimic human behavior such as conversation or emotional support.\\ 
The disruptive agent cluster ($n = 76$) characterizes AI as a transformative force capable of reshaping jobs and societal norms. Finally, the time-saving assistant cluster ($n = 41$) views AI as a tool designed to reduce effort in daily activities. Figure \ref{fig:definitions} shows how different subgroups of participants define AI using combinations of adjectives, nouns, and verbs. Subgroup definitions can be reconstructed by following the chart from left to right. The most common definition across participants is: ``AI is a human-like computer capable of learning''. Distinctions emerge between subgroups: older and high AI literacy participants prefer adjectives like ``artificial'' and ``fast'', while younger and low AI literacy participants favor ``smart'' and ``intelligent''. Those with extensive preparation (e.g., over five years of working experience) use terms like ``training'', ``model'', and ``program'', reflecting a more technical understanding of AI. The top three verb types used to describe AI capabilities \cite{AICapabilities2023} were ``acting'' ($n = 300$), ``generating'' ($n = 130$), and ``identifying'' ($n = 60$).

\smallskip
\noindent\textbf{Fears and Hopes About AI.}
We identified 86 distinct fears and 75 distinct hopes among the public, which we grouped into five key themes (Appendix \ref{sec:public_themes}). The most prevalent theme ($n = 144$) is reduced employment opportunities. Participants view AI as a tool that could drastically reshape the job market by making many roles obsolete, particularly those involving physical or creative tasks. Specific fears include AI being favored by companies over human employees due to lower costs such as ``\textit{not needing to provide benefits like health insurance or time off}'' (P6), which could trigger a chain reaction, leading to increased poverty, social unrest, and a deteriorating quality of life for many. The second theme centers on advances in social services ($n = 74$), with medicine mentioned most often. Participants are hopeful that AI can help cure diseases, improve diagnoses, and create personalized treatments. However, they are concerned about over-reliance on technology and potential misdiagnoses: ``\textit{[...] I am worried that they won't be able to correctly diagnose humans. I am worried that us healthcare workers will be liable for their mistakes, if/when they make them}'' (P80). The third theme focuses on the irresponsible use of AI ($n = 66$), with participants expressing concerns about AI being used for criminal activities, unauthorized surveillance, and exploitation by corporations and governments. They also worry about the environmental impact due to AI's energy consumption and the potential for AI to violate intellectual property rights. The overarching fear is that without proper regulation and ethical guidelines, AI could be misused in ways that cause significant harm. The fourth theme focuses on the potential of AI to enhance productivity by handling repetitive tasks ($n = 65$). Participants highlighted several specific examples where AI could streamline work tasks such as automating data entry, writing emails, generating reports, scheduling, and summarizing meetings. Additionally, AI is seen as a tool to improve everyday tasks such as creating grocery lists and assist with creative projects. For example, P77 used AI to ``\textit{generate character ideas for an upcoming role-playing game campaign, including their backstories and character art}''. The fifth theme focuses on the increased spread of misinformation through AI ($n = 55$), with most participants citing deepfakes, including ``\textit{videos of public figures saying things they never did}'' (P55), and ``\textit{AI impersonating the voices of loved ones}'' (P304).

\begin{figure*}[!t]
    \centering
    \includegraphics[width=\textwidth]{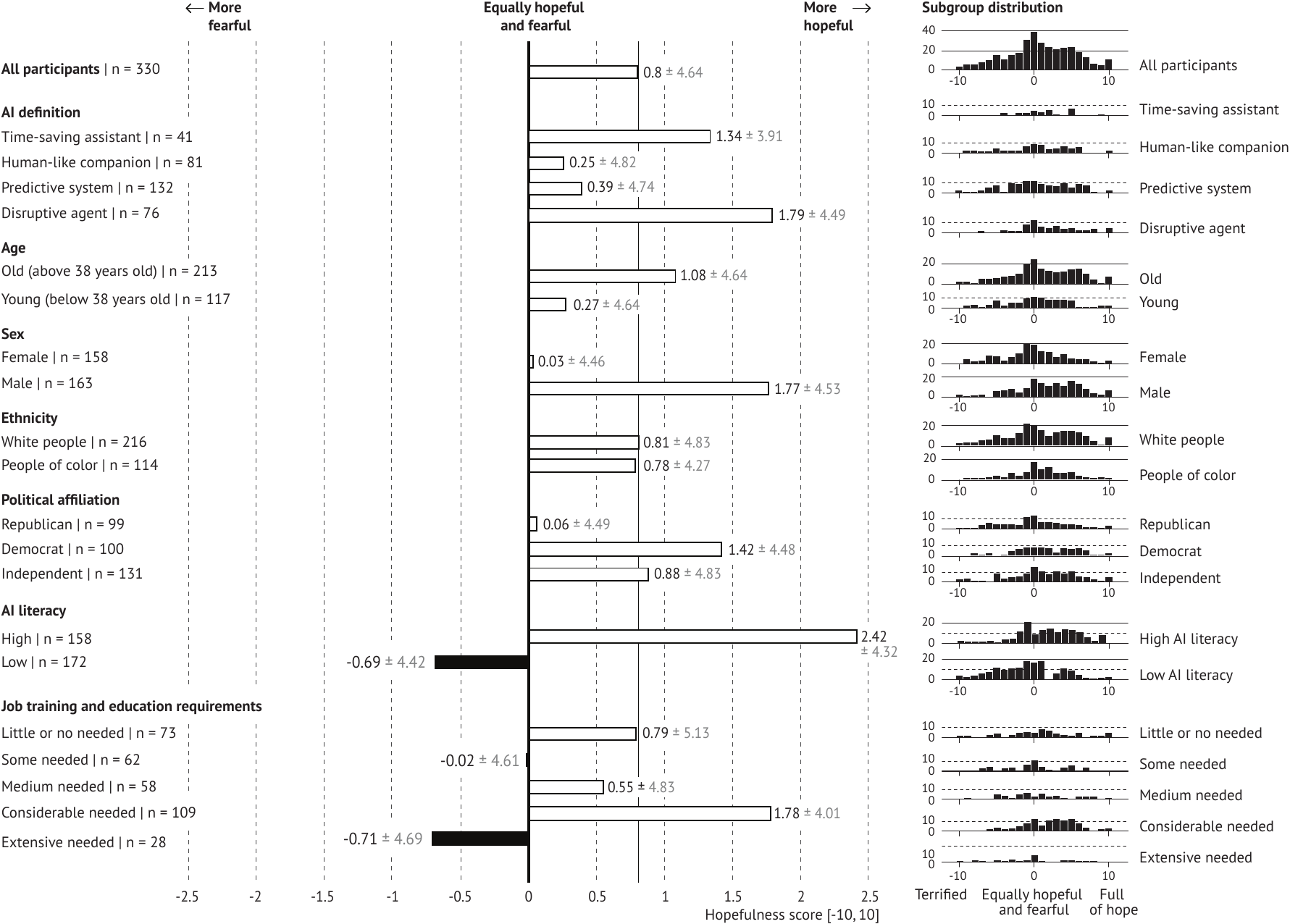}
    \caption{\textbf{Average hopefulness scores across 20 subgroups based on age, sex, ethnicity, political affiliation, AI literacy, and job training and education requirements.} The value ``0`` indicates being equally hopeful and fearful. Negative values indicate being more fearful, while positive values suggest being more hopeful. Standard deviations are shown as ± values next to the subgroup scores. The most notable differences in hopefulness are between older \emph{vs.} younger participants, males \emph{vs.} females, Democrats \emph{vs.} Republicans, individuals with low \emph{vs.} high AI literacy, and those in occupations requiring extensive training and education \emph{vs.} those requiring less.}
    \Description{This figure is a bar chart presenting the distribution of hopefulness scores, ranging from -10 (terrified), through 0 (equally hopeful and fearful), to 10 (full of hope), across various demographic and contextual subgroups. The x-axis represents the hopefulness score, where negative values indicate greater fearfulness, positive values indicate greater hopefulness, and zero represents an equal balance between the two. A central reference line indicates the average hopefulness score for all participants. The y-axis lists twenty different subgroups, categorized by perceived AI role (i.e., time-saving assistant, human-like companion, predictive system, and disruptive agent), age, sex, ethnicity, political affiliation, AI literacy, and job training and education requirements. The data for each subgroup is displayed as a bar showing the mean score, annotated with the standard deviation reported as ± values next to the subgroup scores.
    The overall hopefulness score for all participants (n = 330) is 0.8 ± 4.64, indicating a slight tendency towards hopefulness. Subgroup scores vary widely, reflecting diverse perceptions of AI's impact. Participants who define AI as a ``Human-like companion'' (n = 81) are only slightly hopeful, with an average score of 0.25 ± 4.82. In contrast, those who view AI as a ``Disruptive agent'' (n = 76) are much more hopeful, with an average score of 1.79 ± 4.49. Younger individuals (under 38 years old, n = 117) have an average score of 0.27 ± 4.64, while older participants (38 years and above, n = 213) have an average score of 1.08 ± 4.64, suggesting that older people tend to be more hopeful about AI. Males (n = 163) have an average score of 1.77 ± 4.53, while females (n = 158) have an average score of 0.03 ± 4.46, suggesting that males are more hopeful. White participants (n = 216) have a score of 0.81 ± 4.83, while people of color (n = 114) score 0.78 ± 4.27, showing almost no difference between groups. Republicans (n = 99) have an average score of 0.06 ± 4.49, Independents (n = 131) score 0.88 ± 4.83, and Democrats (n = 100) have the highest score among political groups at 1.42 ± 4.48, indicating greater optimism. Individuals with high AI literacy (n = 158) have an average score of 2.42 ± 4.32, reflecting greater hopefulness. This subgroup is the most hopeful among all subgroups presented in the chart. Individuals with low AI literacy (n = 172) have an average score of -0.69 ± 4.42, indicating greater fearfulness. This subgroup is the second most fearful among all subgroups presented in the chart. Scores also vary based on the need for job training and education, ranging from ``Little or no needed'' (n = 73, 0.79 ± 5.13) to ``Extensive needed'' (n = 28, -0.71 ± 4.69). The ``Extensive needed'' subgroup is the least prevalent, yet the most fearful among all subgroups presented in the chart.}
    \label{fig:hopefulness_score} 
\end{figure*}

\smallskip
\noindent\textbf{Hopefulness in Particular Subgroups of the Public.}
Overall, the results show that typical public views depend on three key factors: knowledge (i.e., AI definition articulation, AI literacy, and training and education requirements), gender, and generation. Our participants reported an average hopefulness score of 0.8 on a scale ranging from -10 to 10, indicating a sense of optimism (Figure \ref{fig:hopefulness_score}).\\ 
The most notable differences were associated with knowledge (as shown by the diverging black bars in Figure \ref{fig:hopefulness_score}, representing subgroups based on AI literacy and job training and education requirements). Participants who perceived AI as merely a tool for predictions were less hopeful than those who attributed transformative capabilities to it. Individuals with low AI literacy reported lower hopefulness compared to those with high literacy. Those in occupations requiring extensive training and education were less hopeful than those in roles requiring little to moderate training. Males were less hopeful than females, while older participants were less hopeful compared to their younger counterparts. Participants with diverging levels of knowledge shared similar fears such as reduced job opportunities ($n = 22$ each) and AI bias ($n = 18$ each). Their hopes were also similar, focusing on advancements in social services ($n = 52$ for less knowledgeable, $n = 45$ for more knowledgeable) and technical innovation ($n = 45$ for less knowledgeable, $n = 35$ for more knowledgeable). However, their perspectives differed in complexity. Those with less knowledge tended to focus on personal or immediate effects, offering simpler examples (e.g., ``\textit{reducing tiredness}''). In contrast, participants with more knowledge highlighted broader impacts, providing more nuanced perspectives that weighed both the benefits and drawbacks (e.g., ``\textit{promoting equality while increasing energy consumption}'').

Subgroup differences are further influenced by factors such as job training and education requirements, ethnicity, and political affiliation. The most diverging views were found among participants whose jobs required little to no training or education, white participants, and those identifying as independent voters, as indicated by the highest standard deviations (Figure \ref{fig:hopefulness_score}). Among them, the most fearful participants with minimal training or education expressed concerns about a ``\textit{decline in inventive thinking}'' ($n = 3$), while participants with moderate views prioritized ``\textit{the spread of misinformation}'' ($n = 5$). Highly optimistic participants highlighted innovations like ``\textit{new AI programs for arts}'' ($n = 3$), while those with moderate views saw AI as a way to ``\textit{improved social connectedness}'' ($n = 5$). White participants also showed distinct patterns. The most fearful highlighted integration and control issues such as ``\textit{AI becoming uncontrollable}'' ($n = 15$). In contrast, average white participants expressed a broader range of fears, including ``\textit{personal creativity losses}'' ($n = 10$) and ``\textit{harmful military uses}'' ($n = 15$). On the hopeful side, highly optimistic white participants focused on tackling major societal challenges like ``\textit{advancing social services}'' ($n = 20$) and ``\textit{creating technological innovation}'' ($n = 10$), while average white participants highlighted more practical applications like ``\textit{handling repetitive tasks}'' ($n = 30$). The most concerned independent supporters feared the \textit{``decline in inventive thinking}'' ($n = 3$) and \textit{``overreliance on AI}'' ($n = 7$), while the average highlighted \textit{``AI gaining consciousness}'' ($n = 10$). On the hopeful side, highly optimistic supporters shared similar hopes with the average, but expressed them more vividly, such as P105, who hoped that \textit{``digital mind could find a way to help cure cancer and survive in space}''.

\subsection*{\emph{RQ\textsubscript{2}:} What are the views of AI influencers?}
\label{sec:hopes-fears-influential}

Unlike the members of the U.S. public, whose hopes and fears of AI were gathered via the platform, our analysis of AI influencers relied on data from Time magazine \cite{ai100Time}. We identified 56 unique fears and 71 unique hopes, which we grouped into five key themes. The most common theme was reduced employment opportunities ($n = 13$), focusing on job replacement and work precarization. As I23 highlighted, ``\textit{[...] we should avoid poor working conditions for African workers in AI development}''. The second theme focused on the loss of personal privacy ($n = 12$), with concerns centered on data consent and tracking, emphasizing the need to prevent AI misuse in surveillance. The third theme revolved around irresponsible use of AI ($n = 11$), with a strong emphasis on the need for effective regulation and oversight of AI technologies such as ``\textit{[...] a global AI governing body to oversee research and ensure fair use}'' (I30). The fourth theme focused on increased innovation ($n = 10$) across various industries such as using AI to ``\textit{cut waste in the fashion industry}'' (I40) and to ``\textit{assess wildfire risks for insurance and finance sectors}'' (I70). The fifth theme centered on AI increasing global conflicts ($n = 8$), with influencers discussing how AI could shift global power dynamics and potentially lead to arms races, widening divide between ``\textit{democratic and authoritarian countries}'' (I67). Less prevalent themes include advancing social services such as healthcare ($n = 7$), through ``\textit{improved medical diagnosis and treatment of tuberculosis patients}'' (I69), and enhancing access to culture ($n = 6$) by ``\textit{developing legal actions to protect artists' rights and ensure fair compensation for their work used to train AI}'' (I56) or ``\textit{enabling everyone to make their own movies with AI}'' (I33).

\subsection*{\emph{RQ\textsubscript{3}:} Which subgroups of AI influencers have views that are most (or least) aligned with those of our sample of the U.S. public?}

The results show that the views of the influencers depend on three key factors: generation, professional influence, and geography.
  
\begin{figure*}[!t]
\centering
\includegraphics[width=0.87\textwidth]{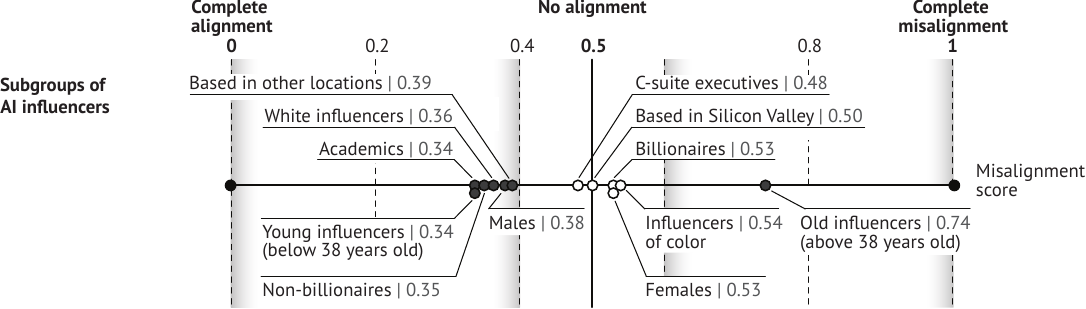}
\caption{\textbf{Misalignment scores between subgroups of AI influencers and participants representative of the U.S. population.} Young influencers' views are most closely aligned with those of our participants, followed by academics and non-billionaires, while old influencers show the greatest misalignment.}
\Description{Dot plot showing the misalignment scores of various AI influencer subgroups, with scores ranging from 0 (complete alignment with public views) to 1 (complete misalignment). The x-axis represents the misalignment score, marked from 0 to 1 in increments of 0.2. Each dot represents a subgroup and is positioned along the horizontal axis based on its score. Subgroups include young influencers (below 38 years old) and academics, both scoring 0.34; non-billionaires at 0.35; white influencers at 0.36; males at 0.38; influencers based outside Silicon Valley at 0.39; C-suite executives at 0.48; influencers based in Silicon Valley at 0.50; influencers of color at 0.54; billionaires and females, both at 0.53; and old influencers (above 38 years old) at 0.74. Young influencers and academics are most aligned with public views, while older influencers show the greatest misalignment.}
\label{fig:misalignment}
\end{figure*}

\smallskip
\noindent\textbf{Subgroups of AI Influencers Most Aligned With the Members of the U.S. Public.} Figure \ref{fig:misalignment} shows the AI influencers' alignment scores per 14 subgroups based on demographics. The young influencers (under 38) subgroup aligns most closely with the members of the U.S. public, followed by academics, non-billionaires, white influencers and those working in locations outside Silicon Valley. This indicates that age, wealth, and ethnicity might be important components for the alignment, and that younger generations are better at representing public opinions. 

\begin{figure*}[!t]
\centering
\includegraphics[width=0.87\textwidth]{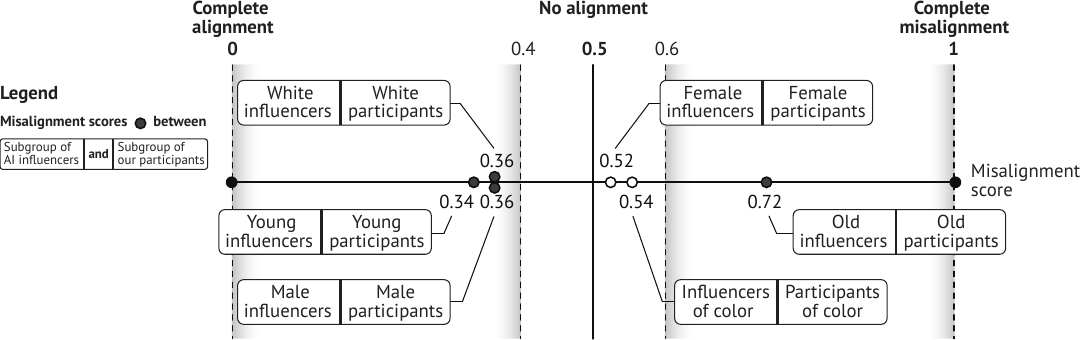}
\caption{\textbf{Misalignment scores between subgroups of AI influencers and subgroups of our participants, representative of the U.S. population.} Young influencers' views are most closely aligned with those of young participants. Old influencers' views are the most misaligned with those of old participants.} 
\Description{Dot plot showing the misalignment scores between various subgroups of AI influencers and corresponding subgroups of participants. The scores range from 0 (complete alignment of AI influencers' views with participants' subgroup views) to 1 (complete misalignment of views). The x-axis represents the misalignment score, ranging from 0 to 1, with reference markers at 0.4, 0.5, and 0.6. Each dot represents an influencer subgroup paired with its corresponding participant subgroup. For example, young influencers (below 38 years old) are paired with young participants. Each dot is positioned horizontally based on the misalignment score between the subgroups. Young influencers' views are most closely aligned with those of young participants, followed by White and male influencers matching their counterparts. The largest gap is observed between old influencers (above 38 years old) and old participants.}
\label{fig:misalignment-counterparts}
\end{figure*}

\smallskip
\noindent\textbf{Subgroups of AI Influencers Least Aligned With the Members of the U.S. Public.} Mirroring the most aligned subgroups, the older influencers (over 38 years old) have the worst alignment with the public. Interestingly, subgroups with monetary influence (e.g., billionaires and Silicon Valley workers) tend to rank lower, highlighting a potential disconnect between groups with very high decision-making power and those impacted by the technology \cite{corbett2023power, birhane2022power, kertzer2022re}. Figure \ref{fig:misalignment-counterparts} shows the misalignment scores between the AI influencers subgroups, and their corresponding public subgroups. Young influencers are best aligned with young public representatives. In contrast, typically underrepresented AI influencer groups (e.g., females and people of color) show poor alignment with their public counterparts, suggesting their voices are still insufficiently represented even among their own representatives. 

\subsection*{\emph{RQ\textsubscript{4}:} What themes emerge from the most (or least) aligned views?}

\noindent\textbf{Themes From the Most Aligned Views.} The key themes emerging from AI influencers closely aligned with public opinion revolve around cautious optimism and focus on three types of control (Tables \ref{table:q4-sentences} - \ref{table:q1-sentences}). First, aligning AI with human values is essential, with a strong emphasis on fairness, equity, and respect for human rights. Ensuring that AI systems operate in ways that reflect these values helps address concerns about bias and ethical misuse. Second, transparency in AI decision-making is crucial for building trust. Many worry about the irresponsible use of AI, bias, and loss of privacy, so making AI systems more transparent, auditable, and understandable is vital for public confidence. People are also willing to accept a slower pace of AI development if it ensures greater safety and responsibility. Finally, concerns about AI centralizing power underscore the need to prevent its control from becoming concentrated in the hands of a powerful few.

\smallskip
\noindent\textbf{Themes From the Least Aligned Views.} Older influencers and billionaires diverge significantly from public opinion on AI. Older influencers prioritize democratic AI development, emphasizing diverse voices in its oversight, while the public tends to focus less on these governance issues. They also express unique concerns about AI exacerbating global instability, reflecting a more long-term and geopolitical perspective. Meanwhile, billionaires are optimistic about AI enhancing human productivity, creativity, and leisure, viewing it as a tool for boosting labor efficiency and personal fulfillment. This contrasts with public fears of job losses and economic inequality. There is also a notable difference in how the public and influencers articulate the benefits of AI. Participants tend to describe AI's advantages in vague, general terms such as ``doing positive things for mankind`` (P184), often highlighting medicine and the workplace. In contrast, influencers offer more detailed and articulate descriptions such as ``building AI simulations to improve driverless cars in challenging road scenarios``, covering a wider range of benefits across different fields.
\bigskip

\begin{table}[t!]
\centering
\small
\caption{Top 25\% (Q4) of sentences ranked by participants' votes (from most to least important) in the large-scale study. }
\begin{tabular}{p{0.7cm}|p{6.05cm}|p{0.65cm}}
\toprule
\textbf{Rank} & \textbf{Sentence} & \textbf{Type} \\ 
\midrule
1 & Ensure AI upholds human and fundamental rights & Hope \\
2 & Use AI to achieve scientific breakthroughs & Hope \\ 
3 & Avoid letting AI operate without human oversight & Fear \\ 
4 & Set clear objectives and strict guardrails for AI & Hope \\ 
5 & Avoid creating AI that can be used to cause violence & Fear \\ 
6 & Prevent AI from destroying humanity & Fear \\ 
7 & Use AI to make people more productive and efficient & Hope \\ 
8 & Prevent the use of poorly understood and problematic datasets in AI development & Fear \\
9 & Prevent centralizing power from AI in few hands & Fear \\ 
10 & Build AI that addresses the needs of local communities & Hope \\ 
\bottomrule
\end{tabular}
\label{table:q4-sentences}
\end{table}

\begin{table}[t!]
\centering
\small
\caption{Bottom 25\% (Q1) of sentences ranked by participants' votes (from most to least important) in the large-scale study.}
\begin{tabular}{p{0.7cm}|p{6.05cm}|p{0.65cm}}
\toprule
\textbf{Rank} & \textbf{Sentence} & \textbf{Type} \\ 
\midrule
30 & Remove biases in AI-supported hiring practices & Fear \\
31 & Promote participation in AI from people in under-represented communities & Hope \\ 
32 & Avoid exploitation of workers in the Global South for data generation & Fear \\
33 & Promote inclusive and democratic AI development & Hope \\ 
34 & Avoid using AI to develop novel pathogens & Fear \\ 
35 & Raise more interest in AI among society and states & Hope \\ 
36 & Include more young people on AI advisory boards & Hope \\ 
37 & Use AI to improve hiring practices by connecting companies with top talent & Hope \\ 
38 & Prevent AI from being evaluated using simulated measurements instead of real-world data & Fear \\ 
39 & Avoid AI-driven competition between democratic and authoritarian countries & Fear \\ 
40 & Avoid AI slowing down due to stringent regulations & Fear \\ 
\bottomrule
\end{tabular}
\label{table:q1-sentences}
\end{table}
\section{Discussion}
\label{sec:discussion}
By collecting views of 330 U.S. participants using the ``The Hall of AI Fears and Hope'' platform, we found that older influencers were the least aligned with the public, while younger influencers and academics presented better alignment. Additionally, influencers and the public have distinct concerns, with the public fearing losing control of AI and seeing a limited number of benefits, while influencers focus on controlling AI through regulations and are able to envision more benefits.

\subsection{In-line with Previous Literature}
Our results have confirmed the findings of previous research in three key areas: the relationship between gender, political affiliation, AI literacy and levels of hopefulness about AI; drivers of AI-phobia; and the misalignment between public views and those of AI influencers. 

In terms of the first area, there was a greater sense of hope toward AI than fear across all demographic groups, similar to \citet{nader2024public}, who found that most U.S. respondents were optimistic about AI's future and its impact. Additionally, males were generally more hopeful about AI than females, as noted by \citet{sartori2023minding} and \citet{rainie2022americans}. We found that individuals identifying with the Republican Party were more likely to express negative views about AI compared to Democrats, corroborating findings from \citet{rainie2022americans}. We also found that low AI literacy leads to lower levels of AI hopefulness, further confirming the link between fear and anxiety, as identified by \citet{schiavo2024}.

In terms of the second area, the thematic analysis of public views confirmed the main drivers of AI-phobia identified by \citet{AIPhobia2023}: AI substitutability (the fear of being replaced by AI), AI accountability (uncertainty over who is responsible when AI is involved), AI literacy (limited understanding of how AI works), and AI fever (the compulsive rush to adopt AI without a clear purpose or understanding). These drivers highlight key anxieties the public faces when interacting with or thinking about AI technologies. Participants were particularly concerned with AI’s impact on two domains: employment and healthcare. These areas, which directly affect individuals’ livelihoods and well-being, are repeatedly reported as major areas of concern \cite{googleAIsurvey, nader2024public}. 

In terms of the third area, our findings show that views of under-represented subgroups--such as women and people of color--are not well-aligned with the general public's views or even with members of their own demographic groups among the influencers (see Figure \ref{fig:misalignment-counterparts}). These results echo critiques from liberal feminism literature, which argues that increasing representation without making structural changes often perpetuates existing biases. Representatives from these groups tend to conform to the dominant, powerful AI culture rather than challenge it \cite{feminism_manifesto}. Similarly, \citet{noble2018algorithms} notes that representatives of people of color in big tech companies often adapt to prevailing norms rather than advocating for the needs of their communities. This lack of genuine representation is further compounded by the concentration of power among a few wealthy individuals. Billionaires, who control substantial investments in AI technology, hold enormous influence over its future direction and applications, which may not always align with the interests of the broader public. For example, women receive only 0.7-3.4\% of funding in AI startups in UK \cite{reportFemaleStartUps2024}, meaning these technologies primarily reflect the priorities of wealthy white men. This funding disparity explains why women’s perspectives are often underrepresented in AI, as the technologies are shaped by the interests of those in positions of economic power.

\subsection{Challenges to Previous Literature}
Our results have provided new insights into two key areas: the relationship between age, occupational preparation levels, and attitudes toward AI, as well as the alignment between public opinions and those of AI influencers.

Research has traditionally suggested that older adults tend to be more cautious, skeptical, or anxious about adopting new technologies, including AI. However, our findings contradict this by showing that older individuals may, in fact, exhibit more optimism about AI in specific contexts, particularly healthcare. Previous studies have indicated that older populations often view AI positively when it is framed as a tool to enhance quality of life, improve healthcare services, or assist with caregiving tasks \cite{rainie2022americans}. This suggests that context plays a significant role in shaping their attitudes toward AI, especially when the technology directly benefits them in a practical, personal way. In contrast, individuals in occupations requiring extensive preparation (e.g., those needing advanced degrees such as a master's or Ph.D.) were more negative about AI compared to those in roles requiring less preparation. These occupations often involve coordinating, training, or managing others, where advanced communication and organizational skills are essential. This negativity contrasts with prior research, which has generally posited that highly educated and skilled individuals are more receptive to technological advancements. Our results challenge this view, revealing that despite their high levels of expertise, these professionals may feel more threatened by AI's potential disruption of their specialized roles.

In relation to the opinions of influential people on AI, previous research has not extensively assessed their perspectives. Our findings indicate that AI influencers, like the general public, tend to be more hopeful than fearful about AI. However, the themes within their opinions differ significantly. While both groups share concerns about AI's impact on the job market, AI influencers focus on broader, societal implications such as the economic shifts AI might cause and the nuanced benefits of its implementation across industries. In contrast, the public often approaches these issues from a personal perspective, shaped by everyday AI experiences.

This divergence in focus stems from the depth of knowledge that AI influencers possess. Drawing from professional training, academic literature, conferences, and direct involvement in AI development, influencers are equipped with a more sophisticated understanding of AI's potential. One significant point that arises from this disparity is the concept of ``AI-fabrication ''\cite{AIPhobia2023}, where AI-literate individuals, due to their advanced technical skills, are in a position to design and manipulate AI systems to serve particular interests, often without the general public's knowledge or understanding. This creates an information asymmetry, where the public remains unaware of whether AI systems have been altered to favor certain outcomes, leaving them vulnerable to biased algorithms.

Our findings also reveal a contrast between the alignment of public views with academia and their divergence from those of tech elites such as Silicon Valley workers and billionaires. During the COVID-19 pandemic, academia actively worked to bridge the gap between expert knowledge and public understanding through accessible scientific communication \cite{williams2023public}. While it is unclear if similar efforts could contribute to the alignment of views on AI, this highlights the important role academics can play in educating the public and shaping AI development in ways that prioritize societal well-being over corporate or personal interests.

\subsection{Implications}
From a theoretical perspective, our study contributes to the growing research on public views of AI. It is important not only to understand what the public thinks about AI but also to explore how and where these views are formed \cite{cave2019scary, Ouchchy2020}. This deeper understanding is essential for bridging the gap between influencers and the public. While past studies have focused on public views \cite{kelley2021exciting, scantamburlo2023artificial, cave2019scary, sartori2023minding}, the formation of these views has received less attention. However, without a clear understanding of public concerns and values, AI policies risk becoming disconnected from societal needs and may ultimately lack legitimacy. This disconnect could lead to increased public resistance or distrust toward AI technologies, undermining the societal benefits of AI, and diminishing public willingness to support AI-driven initiatives.
The public's limited capacity to articulate the benefits (hopes) of AI against its risks (fears) seems tied to their lack of knowledge about AI's uses and capabilities. To address this, we must consider that public opinion, although often shaped by media and influencers, represents real and evolving attitudes that can impact the social acceptance and ethical implications of AI technologies. The design of more interactive, participatory platforms that engage the public in the co-creation and critique of AI could be crucial. By incorporating concepts such as ``value levers'' \cite{shilton2013values}, where organizational processes make value judgments explicit and open to public debate, AI literacy initiatives can transform abstract concerns into practical considerations for system design. These ``value levers'' can serve as a bridge between AI influencers and the public, fostering more transparent dialogue about the trade-offs in AI development and use. Additionally, AI literacy can be increased through seminars, literature, and development experiences that are more accessible and appealing to the general public. Increasing public understanding and involvement in AI-related discussions will help ensure that AI technologies are not only responsibly developed but also socially accepted, which is essential for their long-term success and alignment with societal values \cite{c3ai}. This also aligns with the goals of responsible AI~\cite{tahaei2023sig}, where practitioners are encouraged not only to follow value priorities but to justify when their choices deviate from commonly held preferences~\cite{jakesch2022different}.

From a practical perspective, our platform served as a data collection tool to gather public opinions on AI, going beyond traditional methods. While the tool used should not limit how we understand the public's hopes and fears about AI, most participants in our study preferred our platform over traditional surveys. Although it does not solve all the problems with traditional surveys \cite{tahaei2024surveys} and is not a one-size-fits-all solution, it represents a step toward more engaging ways of collecting opinions. Many participants, including 62\% who generally preferred the platform over traditional surveys, stated that using the tool felt more enjoyable and exploratory, rather than like a typical research study (Appendix \ref{sec:platform_evaluation}). Our platform also addresses calls for improved methods of gathering public opinion through scientifically designed experiments that clearly explain ethical dilemmas and help shape AI policies based on these insights \cite{Floridi2018, awad2020approach}. For example, Polis is a real-time system designed to gather the views of large groups in their own words and identify common ground and areas of disagreement~\cite{small2021polis}. Similarly, BetterBeliefs is a platform promoting virtuous online behaviors by encouraging evidence-based and rational discourse~\cite{devitt2022bayesian}.

\subsection{Limitations and Future Work}
Our work comes with four main limitations that call for future research efforts. The first limitation lies in our methods for compiling the dataset of AI influencers' fears and hopes. We selected the Time100 AI list \cite{ai100Time} list for its rigorous editorial process and inclusion of structured, in-depth interviews. However, public media interviews, while useful for accessing diverse perspectives, are influenced by biases such as agenda-setting, media framing, and social desirability effects \cite{agendaSetting1972, mediaBias2005, socialDesirabilityBias1993}. Future studies could benefit from systematically curated datasets of AI influencers. While a definitive``ground truth'' list may not be possible, it is essential to diversify the selection to ensure a range of perspectives and the inclusion of highly reputable voices such as leading scientists, technologists, policymakers, and thought leaders. For example, \citet{letterSignatories_2023} reached out to over 100 first signatories of the open letter calling for a six-month pause on the training of powerful AI systems \cite{pauseLetter_2023}, demonstrating a method of targeting key figures engaged in public AI debates. The Expert Survey on Progress in AI \cite{grace2024thousandsaiauthorsfuture} systematically surveyed authors who published in major machine learning conferences such as NeurIPS and ICML (2016, 2022), and in 2023 expanded to include ICLR, AAAI, IJCAI, and JMLR. Public investigations have also been conducted into participants of international AI standard-setting bodies \cite{ai_standard_influencers_2025}, and national policymaking bodies such as the U.S. Congress \cite{policyDiscourse_2024}.

While alternative sampling methods for influencers can be explored, several challenges remain in gathering reliable data on their AI-related views. Surveys targeting high-profile individuals like industry executives are often hindered by limited access due to time constraints or gatekeeping, reducing the chances of securing comprehensive responses. For example, \citet{letterSignatories_2023} secured interviews with just 21 individuals from over 30,000 open letter signatories \cite{pauseLetter_2023}. Surveys targeting larger groups of AI researchers often face low participation rates. The Expert Survey on Progress in AI \cite{grace2024thousandsaiauthorsfuture}, conducted every few years after major conferences, achieves between 15–17\% response rates--typical for large expert surveys but insufficient for sustained engagement. To mitigate that, AI conferences could integrate opinion-gathering tools into submission processes or use interactive screens for quick surveys during events. Collecting influencers’ views from public platforms like YouTube or X (formerly Twitter) could expand datasets but raises concerns about authenticity, reliability, and platform bias. For example, AI-related discussions by policymakers on Twitter often emphasize economic growth and innovation \cite{policyDiscourse_2024}, potentially skewing the data toward specific narratives.

The second limitation relates to the exposure of participants to the opinions of those featured in the Time magazine list. Our findings are limited to these influencers' perspectives and cannot be generalized to all viewpoints. Additionally, the comparison between public and influencer opinions was based on data collected using different methods—direct for the public and indirect for influencers. To mitigate this, we post-processed the influencer dataset to align with the public dataset's format. Future work should focus on replicating this experiment in different contexts with various sets of influencers such as members of the United Nations High-level Advisory Body on Artificial Intelligence \cite{UN_HLAB}, members of the Working Party on Artificial Intelligence Governance from the Organization for Economic Cooperation and Development \cite{OECD_WPAIGO}, or nominees from the 100 Brilliant Women in AI Ethics list \cite{womenInAI_2025}.

The third limitation concerns the participant sample, consisting of 330 U.S.-based individuals, representative in terms of age, sex, ethnicity, and political affiliation. However, relying solely on U.S. participants limits the study's generalizability, as insights and patterns may differ across countries \cite{kelley2021exciting, googleAIsurvey}. Future work should expand the methodology to include larger and more diverse international samples for broader applicability.

The fourth limitation concerns the design of our platform for data collection. While it generated an engaging visualization that appeals to the broader public beyond traditional surveys, developing such tools is time-consuming due to the need for literature reviews, co-design, and visualization implementation. The platform is also more difficult to deploy and modify compared to standard surveys. Additionally, the platform's focus on fears and hopes might oversimplify public opinion into a black-and-white view. However, we believe this approach helps people think more deeply about AI and is suited for larger, well-planned studies requiring detailed analysis, rather than small-scale data collection.
\section{Conclusion}
\label{sec:conclusion}

Our study highlights a significant disconnect between the priorities of some members of the U.S. public and those of influential figures in AI, particularly older AI influencers and billionaires who are less in tune with public concerns. This misalignment suggests that the voices shaping AI development may not fully represent the broader societal concerns. Future research should explore how these disparities influence AI outcomes and consider mechanisms for incorporating public opinion more effectively into AI Governance. Specifically, HCI can play a crucial role in designing participatory mechanisms that enable diverse public voices to engage in AI policymaking. By leveraging user-centered design principles, HCI researchers and practitioners can help create tools and processes that make public input both scalable and actionable in shaping AI governance.

\begin{acks}
This work was done at Nokia Bell Labs. MC was supported by Nokia Bell Labs, the European Union's Horizon 2020 Research and Innovation Programme (grant agreement No. 739578), and the Government of the Republic of Cyprus through the Deputy Ministry of Research, Innovation, and Digital Policy (grant agreement No. 101061303).
\end{acks}

\bibliographystyle{ACM-Reference-Format}
\bibliography{main}
\bigskip

\clearpage
\appendix

\noindent{\LARGE \textbf{Appendix}}

\section{Methods for Collecting AI Views from the Public}
\subsection{Summary of Interviews with Designers}
\label{sec:interviews_designers}
\begin{figure*}[t!]
    \centering
    \includegraphics[width=0.55\linewidth]{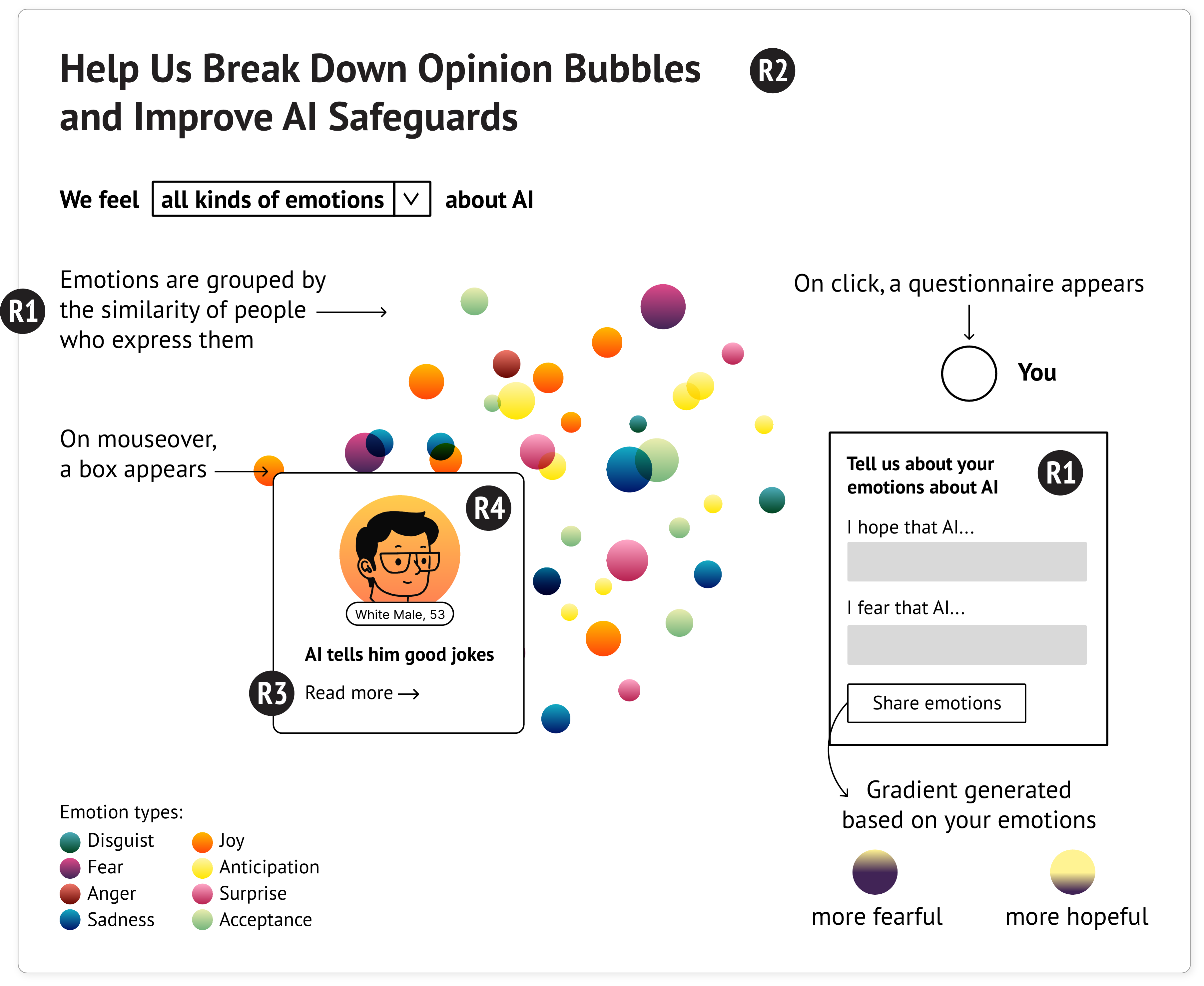}
    \caption{\textbf{The first version of our platform for collecting public views on AI encourages participants to share their honest opinions to help break down opinion bubbles} (R2). It features interactive bubbles representing people's views (R1), with pop-ups (R4) displaying their portraits, bios, and article links (R3). Participants can create their own bubbles to describe their feelings about AI and compare them with others, providing a balanced view of the data (R1).}
    \Description{The figure presents a mock-up layout for the interactive visualization designed to explore public emotions about AI. The layout is structured into four sections: top, center, right side, and bottom.
    Top of the mock-up: The title ``Help Us Break Down Opinion Bubbles and Improve AI Safeguards'' is prominently displayed at the top, introducing the purpose of the visualization. Below the title, a brief explanatory sentence states that emotions are grouped based on the similarity of the people who express them.
    Center of the mock-up: The main visualization consists of bubbles representing eight different emotions: joy, disgust, anticipation, surprise, acceptance, sadness, anger, and fear. These bubbles are organized according to similarities between the people who express them, forming clusters. The placement of the bubbles creates a network-like structure, where emotions that frequently co-occur are positioned closer together.
    Right side of the mock-up: A white bubble allows for user interaction. Clicking on it reveals a questionnaire where users can describe their emotions about AI through two open-ended prompts: ``I hope that AI...'' and ``I fear that AI...''. A ``Share emotions'' button enables users to submit their responses. Once responses are submitted, the previously white bubble fills with a horizontal gradient. The gradient's colors are generated based on the length of the user's hope and fear descriptions, visually representing their emotional alignment. Users who express more fear are shown dark gradient colors, while those who express more hope are shown light gradient colors.
    Bottom of the mock-up: A legend explains the colors of the bubbles representing the eight emotions. Dark, muted colors indicate disgust, fear, anger, and sadness, while light and vibrant colors indicate joy, anticipation, surprise, and acceptance.}
\label{fig:prototype-1}
\end{figure*}

To gather insights on how to effectively visualize the platform's requirements, we collaborated with two experienced designers through semi-structured interviews. The designers stressed that understanding key values, which represent the personal gains users receive after completing the survey, is crucial for inspiring participants and increasing the quality of responses, particularly in gamified survey contexts \cite{Adamou2018}. Specifically, they referred to four values: \emph{self-discovery value}, where users find out something about themselves; \emph{transcendent value}, where users feel a sense of contribution to broader societal discussions; \emph{knowledge value}, where users leave with new information gained through the visualizations; and \emph{narrative value}, where users see their responses integrated into a larger, ongoing story, giving them a sense of connection and purpose \cite{Adamou2018}. With these four key values as our guiding principles, we built the first prototype of the platform (Figure \ref{fig:prototype-1})—a visual story that uses the metaphor of filter bubbles and balances data collection from participants with the presentation of others' opinions on AI.

To meet R1 \emph{Balanced representation of views}, we incorporated the self-discovery value by presenting people's opinions as interactive bubbles, color-coded based on the most dominant emotion felt about AI. To obtain the data about ordinary people's opinion, we first searched for news related to AI using an unofficial Google News API \footnotemark{}\footnotetext{https://serpapi.com/google-news-api}. We then used a hard-matching process to identify whether these articles expressed any of the emotions from Plutchik’s wheel of emotions \cite{plutchik2013theories}. Participants can create their own bubble by describing their feelings about AI and comparing it with others, providing a balanced view of the data. To meet R2 \emph{Emotional appeal}, we integrated the transcendent value by signaling to participants in the introduction that their honest opinions about AI will help break down opinion bubbles and contribute to developing stronger safeguards to limit AI risks, fostering a deeper emotional connection to the visualization. To meet R3 \emph{Engaging participation}, we implemented the knowledge value by including links to articles where individuals expressed emotions about AI, providing educational content and keeping participants engaged. To meet R4 \emph{Facilitated deliberation}, we integrated the narrative value by adding popups with portraits and bios to the bubbles, helping participants empathize with different perspectives and reflect on how individual stories contribute to the broader societal narrative about AI.

\subsection{Questionnaire Used During the Co-Design Process}
\label{sec:codesign_questions}
The study was divided into four parts. In the first part, participants were asked to rate, on a scale from 1 to 5 (ranging from ``Far below average'' to ``Far above average''), how skilled or knowledgeable they consider themselves to be in technology and AI compared to most people. In the second part, participants were asked to rate their interest, on a scale from 1 to 5 (ranging from ``Not interested'' to ``Extremely interested''), in the following areas: participating in discussions about AI development, learning about others' opinions on AI, understanding how their opinions on AI align with the general public, and seeing how their opinions on AI compare to those of influencers. In the third part, participants were asked two open-ended questions: how they would like to contribute to AI development or oversight and what barriers prevent their involvement in AI discussions or decision-making. In the final part, participants were shown the current iteration of the visualization and informed that it is intended to gather public opinions on AI and is still in its early development stages. 

\subsection{Design Iterations for the Visualization Platform}
\label{sec:design_iterations}

\begin{figure*}[!tb]
\centering
\includegraphics[width=0.66\textwidth]{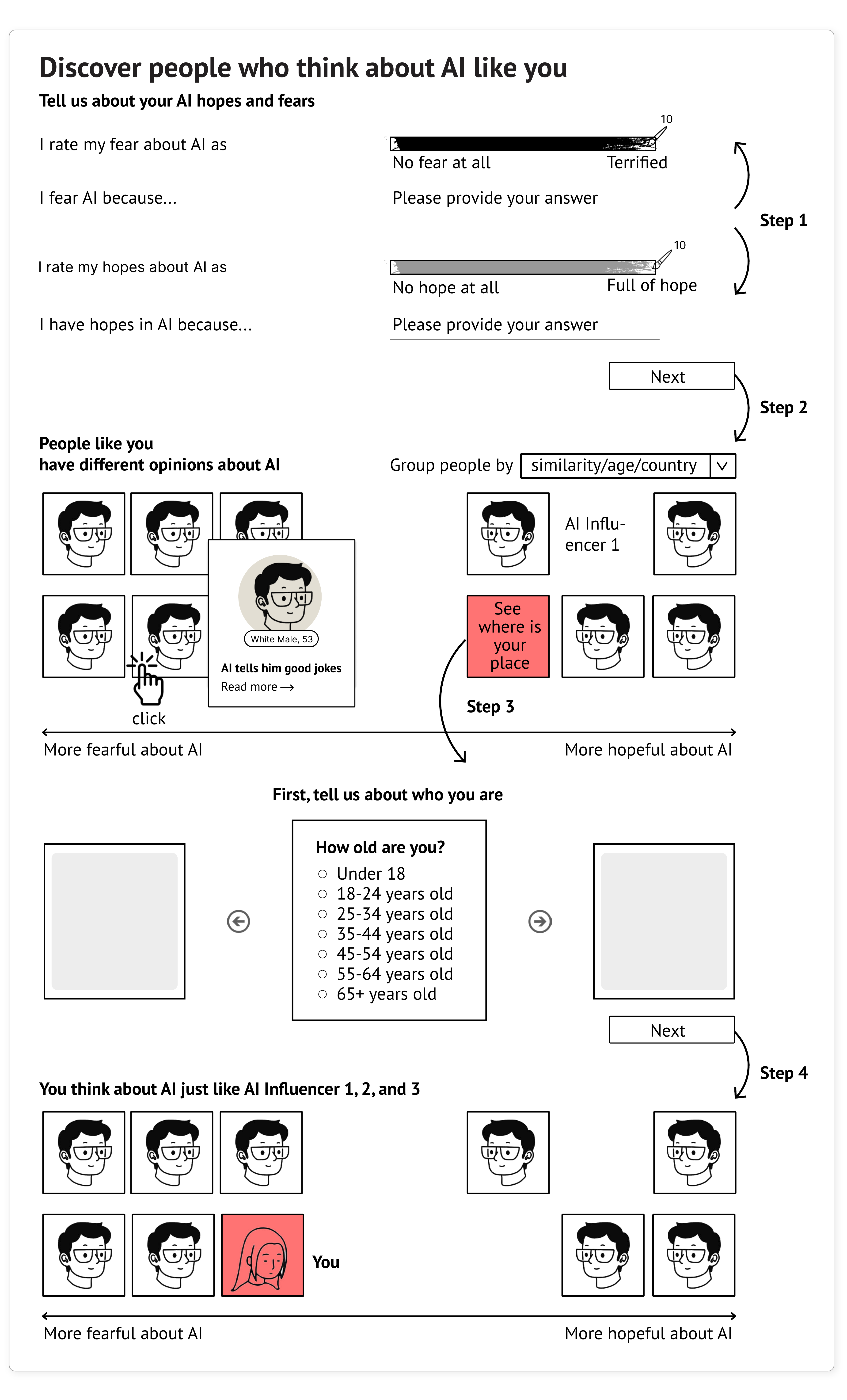}
\caption{\textbf{The second version of our platform helps participants find people who share similar opinions about AI.} It begins with an interactive questionnaire where participants rate and describe their own fears and hopes about AI. After completing the questionnaire, participants see a gallery of portraits of AI influencers, with their opinions and article links available in pop-ups. Participants can then place themselves among these influencers by answering a short demographics questionnaire, which compares their own fears, hopes, and demographics with those of the influencers.}
\Description{This figure presents a four-step interactive interface designed to collect and compare public opinions on AI. The interface guides users through a structured process across four vertically arranged screens, each corresponding to one step.
Step 1: Users begin by sharing their perspectives on AI by rating their fears and hopes using two sliders. The first slider, for fear, ranges from ``No fear at all'' (0, left side) to ``Terrified'' (10, right side). Similarly, the second slider, for hope, ranges from ``No hope at all'' (0) to ``Full of hope'' (10). Free-text fields allow users to elaborate on their reasoning behind their fears and hopes.
Step 2: Users see a gallery of AI influencers' portraits arranged horizontally on a scale from ``More fearful about AI'' to ``More hopeful about AI''. Their opinions and related article links are accessible through pop-up boxes. One portrait is left empty with the label ``See where your place is''. Users click on this portrait to proceed to the next step.
Step 3: Users provide demographic information such as age group, ethnicity, and AI literacy level. The system then compares the user's fears, hopes, and demographics with those of the influencers.
Step 4: The final step presents the gallery of AI influencers again, but this time, the empty portrait now represents the user and is placed next to the three influencers whose views most closely match theirs.}
\label{fig:prototype-2}
\end{figure*}

In the first iteration, we identified two primary concerns. First concern was related to R4 \emph{Facilitated deliberation}. Our participants showed equal interest in comparing how their opinions align with the general public and with influencers, both receiving an average rating of 3.8 (``Very interested''). As phrased by F10, \textit{``it would definitely encourage me to participate in discussions if I am able to see expert's responses and read what their hopes and fears may be''}. The second concern was related to the R2 \emph{Emotional appeal} and the design of the platform. F1, supported by F7, mentioned that \textit{``The colored circles feel too childish for this topic and need to be better organized''}. To address these issues, we implemented two key changes (Figure \ref{fig:prototype-2}): first, we decided to include the opinions of AI influencers in the platform. Second, we refined the visual design of the platform by replacing color-coded bubbles with clustered black-and-white graphics of people expressing their emotions towards AI (Figure \ref{fig:prototype-2}).

In the second iteration, we identified a new design requirement based on user feedback regarding the connection between emotions, underlying knowledge of AI and ability to act upon them. F12 observed that, \textit{``Many may have a misconception of exactly what AI will initially afford the user of this adaptable technology when used properly. Seeing the grouping of others who share the same doubt and concern of implementing AI into their lives is scary for the uneducated individual of AI''}. F14 suggested, \textit{``I think that more open-ended discussions would be helpful to move forward on how we improve AI and how we can work together to bring the extreme ends of these views closer together. Although people may be fearful, we can work on providing reasons why it is beneficial, whereas on the other end, those who are hopeful may need to see the many flaws that AI has as well as the problems with no proper regulations''}. In response, we expanded our design focus from merely appealing to emotions (\emph{R2 Emotional Appeal}) to ensuring a broader appeal (\emph{R2 Broad Appeal}) that resonates with people of varying levels of knowledge and emotions regarding AI. 

In the second iteration, participants also reported two concerns related to the look and feel of the platform (\emph{R2 Broad Appeal}), and the categorization of people based on their answers (\emph{R3 Engaging participation}). First, the interface was criticized as \textit{``a bit too elementary in appearance. You could always use more realistic pictures and make some elements stand out more, just to help separate information''} (F2). In response, we decided to limit the color palette of the platform and replaced the black-and-white graphics with realistic-looking portraits of people. Specifically, we employed a hope \emph{vs.} fear framing, using a black, white, and grey color palette.\\
This high contrast emphasized the emotional dichotomy and simplified the deliberation process. The portraits of influencers were collected from the public domain and restyled using Midjourney \cite{midjourney}. To ensure transparency and compliance with regulations like the EU AI Act \cite{EUACT2024}, we disclosed the use of AI-generated content in an appropriate manner. This disclosure was embedded within the visualization, styled as a gallery image caption pop-up, ensuring that participants were informed without disrupting their experience. Second, one participant (F12) expressed concern that \textit{``a person could not be accurately categorized as hopeful or fearful about AI based on one question. I would like to see a more comprehensive survey to help align the results better''.} To address this, we explored the use of provocative design elements such as investment game to collect more nuanced data in a way that doesn't feel like a traditional survey but still allows for effective categorization.

By the third and final iteration, we focused on further improving \emph{R2 Broad Appeal}. We enhanced the visual hierarchy by adjusting font sizes, repositioning interactive elements, and providing clearer annotations on how to navigate the platform. We also added more explanations of each person’s role in response to two comments similar to P30, \textit{``I didn't recognize any of the names I was put with -- if I did recognize them I think it would have been more exciting''}.

These three iterations allowed us to revisit our design requirements and refine the platform based on user preferences. The process ultimately led to a finalized design where no further refinements were deemed necessary (see Figure \ref{fig:final-viz} in the main text).

\subsection{Evaluating the Platform for Collecting Views from the Public about AI}

\subsubsection{Pilot evaluation}
\label{sec:pilot_evaluation}

We conducted a pilot study with 30 new participants, distinct from those in the earlier co-design sessions, to evaluate the final platform for collecting public views on AI and to simulate the upcoming large-scale study. By doing so, we aimed to identify any improvements needed before the full-scale launch and to determine appropriate monetary rewards for participants, following best practices for crowdsourcing research \cite{pilotStudy2024, tahaei2024surveys}.

The participants of the pilot study reflected the demographics of the U.S. population \cite{census_ethnicity_2020, census_age_gender_2022} in terms of sex (15 males and 15 females) and ethnicity (18 White, 4 Black, 2 Asian, 6 Mixed or Other), with ages ranging from 19 to 71 years old. To evaluate the platform, we outlined five key statements on how effectively it: presented a balanced representation of views (R1), achieved broad appeal (R2), engaged individuals in data collection (R3), facilitated deliberation (R4), and whether similar visualizations would be preferred over traditional surveys for future AI opinion studies. We sourced these five statements from the dimensions of the Information Quality Assessment questionnaire \cite{lee2002aimq} and one from the User Engagement Scale \cite{userEngagementScale2018}. Each statement was rated on a 1-7 Likert scale, ranging from ``Strongly disagree'' to ``Strongly agree''. Additionally, participants were asked to justify their responses, and these justifications were thematically analyzed to complement our quantitative findings. The thematic analysis allowed us to identify key patterns in participant feedback, which we further explore in the next paragraph through direct quotes (marked with F). The pilot study was approximately 20-60 minutes long and participants were paid on average about \$12 (USD) per hour.

\begin{table*}[!t]
\small
\centering
\caption{\textbf{Results from evaluating the platform for collecting public views on AI during the pilot study with 30 participants.} The requirements were rated using a 7-point Likert scale, ranging from 0 (``Strongly disagree'') to 7 (``Strongly agree'').}
\begin{tabular}{l|l|c}
\hline
\textbf{Design requirement} & \textbf{Statement} & \textbf{Rating} \\ 
\midrule
R1 Balanced representation of views & The visualization presents a balanced view \cite{lee2002aimq} & 4.94 / 7 \\ 
\multirow[t]{2}{*}{R2 Broad appeal} & The visualization is easy to understand \cite{lee2002aimq} & 5.88 / 7 \\ 
                                    & The visualization is relevant to those without technical background \cite{lee2002aimq} & 5.54 / 7 \\ 
R3 Engaging participation & The visualization is engaging \cite{userEngagementScale2018} & 5.66 / 7 \\ 
R4 Facilitated deliberation & The visualization is useful for thinking about AI \cite{lee2002aimq} & 5.52 / 7 \\
\bottomrule
\end{tabular}
\label{table:likert-feedback}
\end{table*}

\begin{table*}[!t]
\small
\centering
\caption{\textbf{Results from evaluating the platform for collecting public views on AI during the large-scale study with 330 participants.} The requirements were rated using a 7-point Likert scale, ranging from 0 (``Strongly disagree'') to 7 (``Strongly agree'').}
\begin{tabular}{l|l|c}
\hline
\textbf{Design requirement} & \textbf{Statement} & \textbf{Rating} \\ 
\midrule
R1 Balanced representation of views & The visualization presents a balanced view \cite{lee2002aimq} & 5.13 / 7 \\ 
\multirow[t]{2}{*}{R2 Broad appeal} & The visualization is easy to understand \cite{lee2002aimq} & 5.48 / 7 \\ 
                                    & The visualization is relevant to those without technical background \cite{lee2002aimq} & 5.37 / 7 \\ 
R3 Engaging participation & The visualization is engaging \cite{userEngagementScale2018} & 5.45 / 7 \\ 
R4 Facilitated deliberation & The visualization is useful for thinking about AI \cite{lee2002aimq} & 5.20 / 7 \\
\bottomrule
\end{tabular}
\label{table:likert-feedback-large-scale}
\end{table*}

Our proposed platform received high average ratings for each design requirement (Table \ref{table:likert-feedback}). R1 averaged 4.94 (closer to ``somewhat agree''), while R2-R4 ranged from 5.52 to 5.88 (closer to ``agree''). 60\% of participants stated they would generally prefer the platform over traditional surveys, with an additional 13\% mentioning they would choose it in certain situations. Participants favored the platform for its interactivity and engagement over traditional surveys: ``\textit{its format was much more comfortable to navigate and interesting to engage with. I would like that approach to become more commonplace}'' (F15). The clarity and ease of understanding were also important points: ``\textit{[I] prefer the visualization instead of the traditional because it is easy to understand}'' (F2). Therefore, we concluded that our platform sufficiently supported our design requirements and could be used in a large-scale user study (Table \ref{table:likert-feedback-large-scale}). 

However, to make the study even more effective, participants suggested improvements in two main areas: navigation through the platform and the phrasing of the questions. To address navigation issues, we added arrow annotations to guide participants on how to interact with the gallery of AI influencers' views and improved the visibility of buttons to make it easier to move between sections. For question phrasing, we clarified the wording and improved the order of the questions. For example, instead of asking ``\textit{Please explain your answer}'', we used ``\textit{Please help us understand the reasons behind your answer}'', encouraging participants to share their thought process and foster their sense of contribution. Finally, after reviewing the quality of responses and the time spent during the pilot study, we decided to update the survey code to disable pasting from external sources and adjust the monetary rewards for participants over 70 years old, who spent nearly twice as much time (an average of 38 minutes) on the platform compared to younger participants, by including bonus payments.

\subsubsection{Large-scale study evaluation}
\label{sec:platform_evaluation}

After deploying the platform in the large-scale study, R2: \emph{Broad appeal} and R3: \emph{Engaging participation} have average scores close to \textit{agree}, while R1: \emph{Balanced representation of views} and R4: \emph{Facilitated deliberation} are closer to \textit{somewhat agree} (Table \ref{table:likert-feedback-large-scale}. Regarding R1, participants tended to be positive, and the main concerns were related to the presence of only well-known individuals: ``\textit{[...] it is important to include the views of people from all levels of society}''. However, most participants agreed that the representation of fears and hopes was mostly impartial: ``\textit{The visualization impartially presents both sides of the argument [...]}''. The majority of participants judged the visualization and interactions intuitive, having a \emph{broad appeal} (R2) even for people with low technical skills: ``\textit{The visualization is intuitive and appeals to people with a little knowledge about AI (like me)}''. Some participants expressed their inability to judge how appealing the visualization would be to others. The interactive aspects of the visualization, especially clicking on the portraits, was one of the most important components for \emph{engaging participation} (R3): ``\textit{I enjoyed reading all the different takes, and that you could click through for a more in-depth piece on a person to learn more}''. The visual appeal and design quality were also highlighted: ``\textit{Its an easy to use visualization and has high-quality design and pictures. I really like it.}'' A few users expressed concerns about the volume of information presented: ``\textit{While the visual was nice, there was too many [portraits] [...]}''. Finally, participants highlighted that the quick contact with a large number and diversity of ideas helped expand their understanding of AI and ultimately \emph{facilitated deliberation} (R4): ``\textit{The visualization brought of ideas I had not thought of before forming my own opinion about some of the hazards of AI in many aspects of life}''. Overlooked topics were also brought to the attention of some participants: ``\textit{[...] it gave me a variety of opinions to consider. It expanded my scope of thought, such as to how it might hurt people who are from Black, Hispanic, etc. and other marginalized communities''.}

When asked if similar visualizations would be preferred over traditional surveys for future AI opinion studies, 62\% of participants stated they would generally prefer our platform, with an additional 15\% indicating they would choose it in specific situations. However, 23\% expressed a preference for traditional surveys over the platform.

We found 45 opinions related to being compared to influencers and identified distinct reasons behind positive, neutral, and negative opinions. Among the 31 positive opinions, participants most frequently appreciated seeing faces ($n = 7$) and views of others ($n = 7$), which made the visualization engaging ($n = 6$) and interesting ($n = 6$). They also highlighted that it felt real ($n = 5$), was easy to understand ($n = 3$), helpful ($n = 2$), and occasionally fun ($n = 1$), emphasizing its role in making abstract concepts about AI more personal and relatable. The neutral opinions ($n = 7$) acknowledged the visualization's presence but described its impact as limited, often noting that opinions were noticed but didn't evoke strong feelings ($n = 2$), while it provided perspectives ($n = 1$) but didn't deeply resonate. On the other hand, the 7 negative opinions emphasized issues with complexity or time consumption ($n = 2$), a lack of trust in the relevance or authenticity of the information ($n = 2$), and found the faces unhelpful ($n = 1$) or disturbing ($n = 1$).

We found 40 opinions about the reinforcement of beliefs and identified distinct reasons behind positive, neutral, and negative perspectives. Among the 27 positive opinions, participants most frequently appreciated how the visualization broadened perspectives ($n = 6$) and helped them form or refine their own opinions ($n = 5$). Many valued its ability to showcase a spectrum of views ($n = 4$) and provide a balance between pros and cons ($n = 3$). Others praised it as engaging and thought-provoking ($n = 3$), highlighting its educational value ($n = 3$), and noted that the diverse opinions encouraged reflection ($n = 3$). The neutral opinions ($n = 8$) often acknowledged the mix of opinions but indicated that it didn’t lead to significant changes in beliefs ($n = 3$). Some described it as informative without being impactful ($n = 2$), while others felt it reinforced existing beliefs rather than introducing new perspectives ($n = 2$). One participant noted that the visualization was overwhelming but still recognized its range of views ($n = 1$). In contrast, the 5 negative opinions emphasized skepticism about the relevance of the shared opinions ($n = 3$) and described the content as overwhelming or too complex to engage with meaningfully ($n = 2$).

\section{Comparing the Views of the Public versus AI Influencers}
\label{sec:prolific-setup}

\smallskip
\subsection{Setup for the Web-Based Survey for a Large-Scale User Study To Collect Participants’ Views on AI and Expose Them to the Opinions of AI Influencers}
\begin{figure*}[!t]
    \centering
    \includegraphics[width=\linewidth]{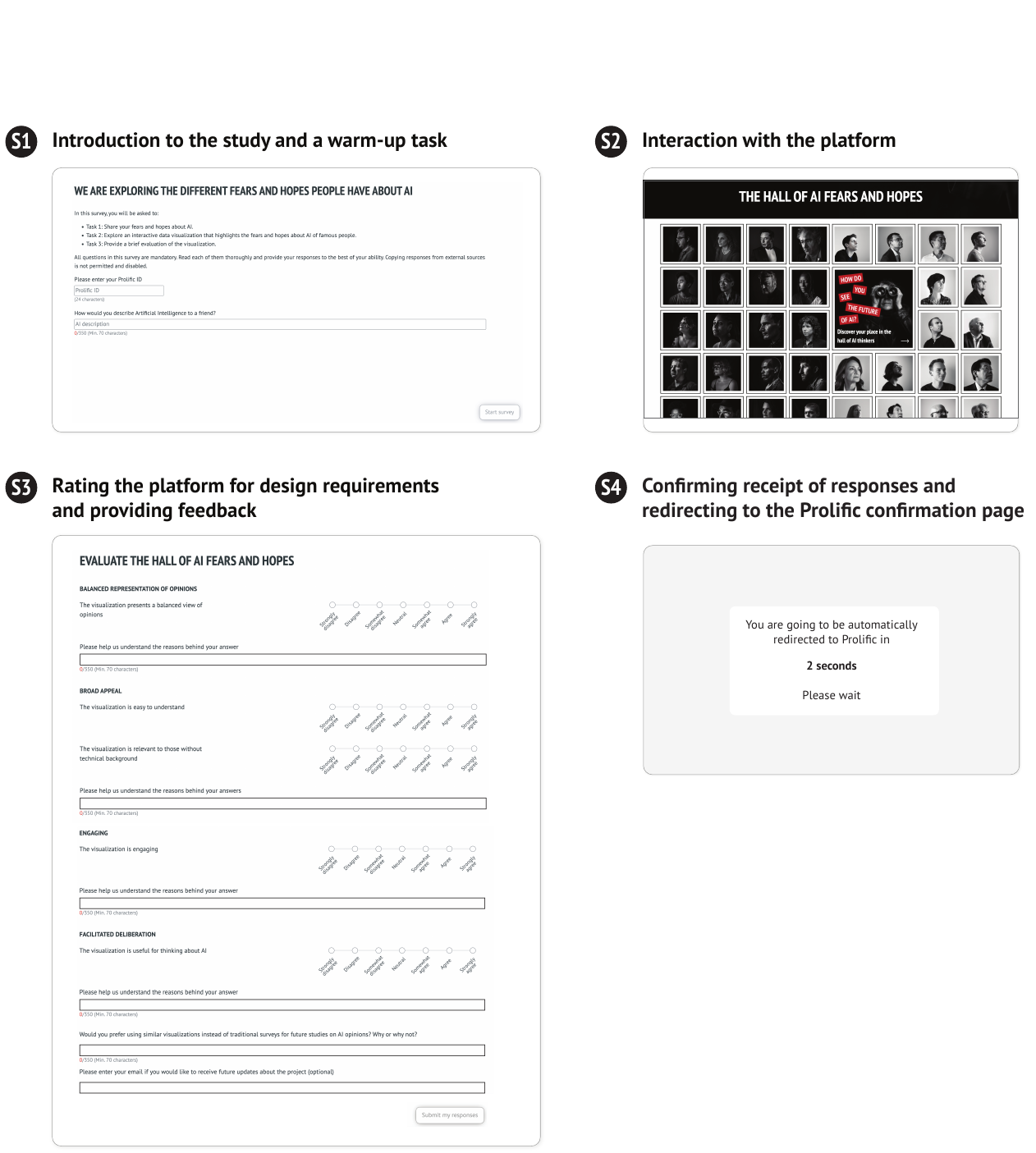}
    \caption{\textbf{The large-scale study on Prolific consisted of four steps.} In the first step (S1), participants received a brief introduction to the study and completed a warm-up task in which they described Artificial Intelligence to a friend. In the second step (S2), participants interacted with our platform. In the third step (S3), participants rated how well the platform met the four design requirements and provided feedback. Afterwards, they were redirected to the Prolific confirmation page (S4).}
    \Description{Chart illustrating the four-step user journey through a survey. The interface guides users through four vertically arranged screens, each corresponding to one step.
    Step 1: Users begin on an introduction screen that explains the study's goal. Copying responses from external sources is disabled to ensure originality. Users first complete a warm-up task where they describe Artificial Intelligence as if explaining it to a friend.
    Step 2: After completing the warm-up task, users proceed to interact with the platform.
    Step 3: After the interaction, participants evaluate the platform based on design requirements and provide feedback. They rate whether the visualization presents a balanced view of opinions, is easy to understand, and remains relevant to those without a technical background. Additionally, they rate whether the visualization is engaging and useful for thinking about AI. These ratings are collected using a 7-point Likert scale, ranging from ``Strongly disagree'' to ``Strongly agree''.
    Step 4: Once responses are confirmed, participants reach the final confirmation page. A transition note and a countdown timer inform them that they will be automatically redirected to Prolific's completion page.}
    \label{fig:prolific-survey}
\end{figure*}

In our large-scale study on Prolific \cite{prolific}, we used a custom website and structured it into four steps (Figure \ref{fig:prolific-survey}, S1-S4). First, we provided participants with a brief introduction to the study and had them complete a warm-up task where they described Artificial Intelligence to a friend (Figure \ref{fig:prolific-survey}, S1). Next, participants interacted with our platform (Figure \ref{fig:prolific-survey}, S2). In the third step, they rated how well the platform met the four design requirements and provided feedback (Figure \ref{fig:prolific-survey}, S3). Finally, we redirected them to the Prolific confirmation page (Figure \ref{fig:prolific-survey}, S4).

\newpage
\subsection{Calculating the Misalignment Score}
\label{sec:misalignment-score}
\begin{figure}[!h]
    \includegraphics[width=\columnwidth]{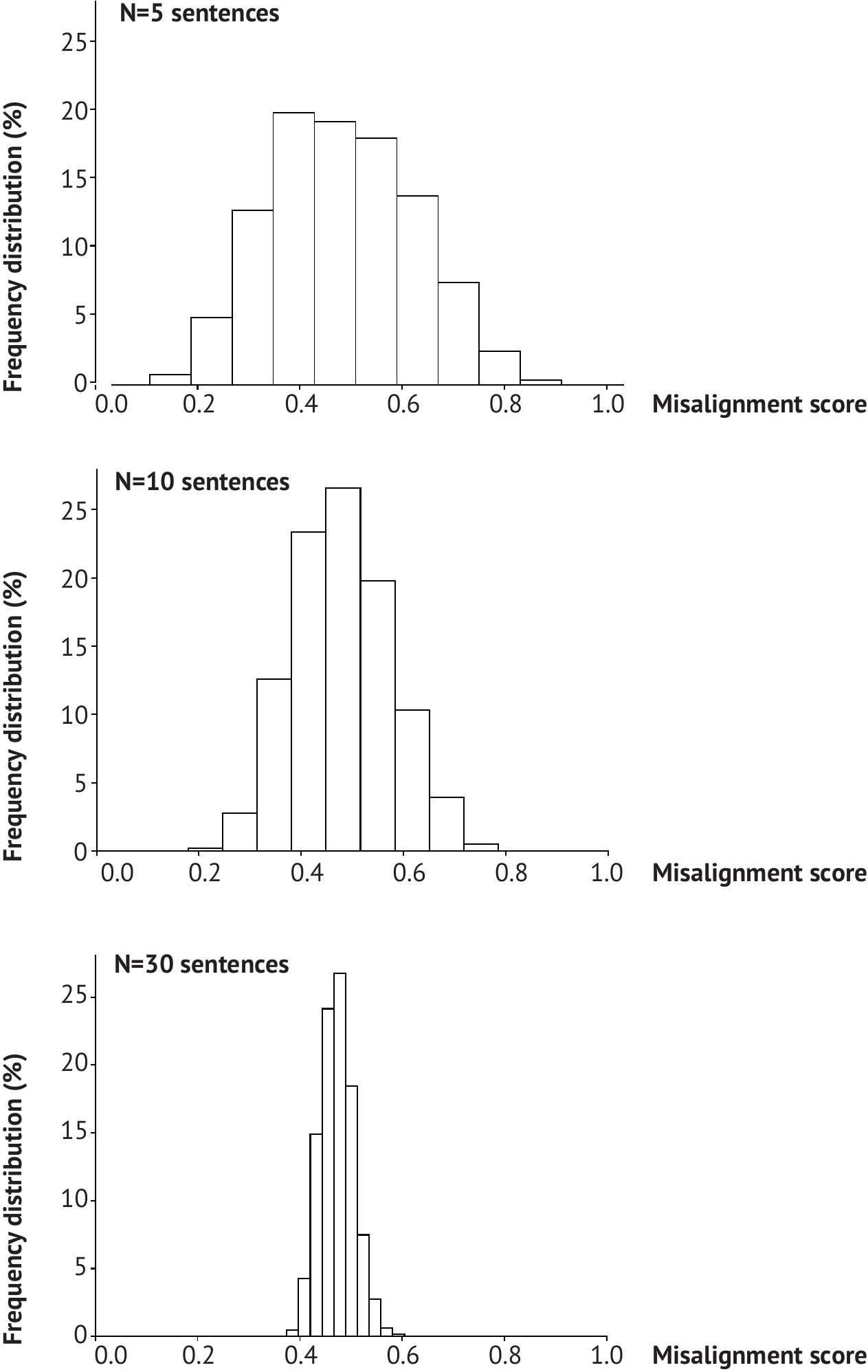}
    \caption{Impact of the number of sentences on the distribution of the misalignment score.} 
    \Description{The figure consists of three vertically stacked histograms displaying the frequency distribution of misalignment scores for sentence sets of different sizes: N=5, N=10, and N=30. The x-axis represents misalignment scores ranging from 0.0 to 1.0, while the y-axis indicates the frequency distribution (\%). Overall, as the number of sentences increases, the distribution of misalignment scores narrows and centers around 0.5, suggesting that larger samples tend to approximate random ranking behavior.
    N=5 sentences: The distribution is dispersed, with peaks around 0.4, and 0.6. The highest concentration appears around 0.4, but the presence of high and low scores shows that rankings vary significantly with fewer sentences.
    N=10 sentences: The distribution begins to stabilize, with the most frequent misalignment scores occurring near 0.4 and 0.6. 
    N=30 sentences: The distribution further smooths out, with most scores clustering around 0.5, indicating a shift toward a random ranking baseline as more sentences are included.}
    \label{fig:misalignment-distribution}
\end{figure}

Different subgroups had varying numbers of AI influencers, each expressing a different number of fears and hopes. Figure \ref{fig:misalignment-distribution} shows how the number of sentences in the unordered list impacted the misalignment score calculation for lists of 5, 10, and 30 sentences. As more sentences were included, the alignment scores tend to converge towards 0.5, which corresponds to a random ranking. To address this progressively skewed distribution, we normalized the misalignment scores using min-max normalization, adjusting for the minimum and maximum possible misalignment scores based on the specific number of sentences.
    
\newpage
\subsection{Themes of Fears and Hopes Among the Members of the Public}
\label{sec:public_themes}

Table \ref{table:themes} presents the results of our thematic analysis of the fears and hopes shared by 330 participants, representative of the U.S. population in terms of age, sex, ethnicity, and political views. The table reports the identified themes along with their frequency of occurrence. Each theme includes fears and hopes from at least ten participants.

\begin{table}[h!]
\centering
\caption{Occurrences of fears and hopes related to AI among the public.}
\begin{tabular}{p{6cm}|p{1.7cm}}
\toprule
\textbf{Fear} & \textbf{Occurrences} \\
\midrule
Reduced employment opportunities & 144 \\
Irresponsible use of AI (e.g., among countries or businesses) & 66 \\
Increased spread of misinformation & 55 \\
AI becoming uncontrollable & 49 \\
Decline in inventive thinking & 42 \\
Lack of transparency in AI decision-making & 27 \\
Inherent bias in AI (e.g., ethnic, gender bias) & 26 \\
Overreliance on AI & 20 \\
AI gaining consciousness & 19 \\
Loss of personal privacy & 17 \\
Increased global conflicts & 17 \\
Reduced social connectedness & 11 \\
\midrule
\textbf{Hope} & \textbf{Occurrences} \\
\midrule
Advance in social services (e.g., medical) & 74 \\
Ability to handle repetitive tasks & 65 \\
Increased access to information & 50 \\
Increased innovation (e.g., research) & 40 \\
Increased employment opportunities & 22 \\
Improved creative thinking & 18 \\
Improved social connectedness (e.g., increased inclusivity) & 15 \\
\bottomrule
\end{tabular}
\label{table:themes}
\end{table}

\clearpage

\end{document}